\documentclass[amsfonts,amsmath,amssymb,prl,twocolumn,superscriptaddress]{revtex4-2}

\usepackage[compact]{titlesec}
\usepackage[utf8]{inputenc}
\usepackage{graphicx}
\usepackage{dcolumn}
\usepackage{bm}
\usepackage{natbib}
\usepackage{float}
\usepackage{amsmath}
\usepackage[dvipsnames]{xcolor}
\usepackage{midfloat}
\usepackage{datetime2}
\usepackage{ragged2e}
\usepackage{lineno}
\usepackage{indentfirst}
\usepackage{gensymb}
\usepackage{comment}

\makeatletter

\def\maketitle{
    \@author
    \title@column
    \titleblock@produce
    \suppressfloats[t]
}

\def\makesuptitle{
    \centering{\textbf{\large{Supplementary Materials for\\}}}
    \@author
    \titleblock@produce
    \suppressfloats[t]
}

\makeatother

\newcommand{\beginsupplement}{
    \clearpage
    \onecolumngrid
    \makesuptitle
    \setcounter{table}{0}
    \renewcommand{\thetable}{S\arabic{table}}
    \setcounter{figure}{0}
    \renewcommand{\thefigure}{S\arabic{figure}}
    \setcounter{equation}{0}
    \renewcommand{\theequation}{S\arabic{equation}}
    \setcounter{secnumdepth}{0}
}

\begin{document}

\title{Torsional Force Microscopy of Van der Waals Moir\'es and Atomic Lattices}

\author{Mihir Pendharkar}
\email{mihirpen [at] stanford [dot] edu}
\affiliation{Stanford Institute for Materials and Energy Sciences, SLAC National Accelerator Laboratory, Menlo Park, CA 94025}
\affiliation{Department of Materials Science and Engineering, Stanford University, Stanford, CA 94305}

\author{Steven J. Tran}
\affiliation{Stanford Institute for Materials and Energy Sciences, SLAC National Accelerator Laboratory, Menlo Park, CA 94025}
\affiliation{Department of Physics, Stanford University, Stanford, CA 94305}

\author{Gregory Zaborski Jr.}
\affiliation{Stanford Institute for Materials and Energy Sciences, SLAC National Accelerator Laboratory, Menlo Park, CA 94025}
\affiliation{Department of Materials Science and Engineering, Stanford University, Stanford, CA 94305}

\author{Joe Finney}
\affiliation{Stanford Institute for Materials and Energy Sciences, SLAC National Accelerator Laboratory, Menlo Park, CA 94025}
\affiliation{Department of Physics, Stanford University, Stanford, CA 94305}

\author{\\Aaron L. Sharpe}
\affiliation{Materials Physics Department, Sandia National Laboratories, Livermore, CA 94550}

\author{Rupini V. Kamat}
\affiliation{Stanford Institute for Materials and Energy Sciences, SLAC National Accelerator Laboratory, Menlo Park, CA 94025}
\affiliation{Department of Physics, Stanford University, Stanford, CA 94305}

\author{Sandesh S. Kalantre}
\affiliation{Stanford Institute for Materials and Energy Sciences, SLAC National Accelerator Laboratory, Menlo Park, CA 94025}
\affiliation{Department of Physics, Stanford University, Stanford, CA 94305}

\author{Marisa Hocking}
\affiliation{Stanford Institute for Materials and Energy Sciences, SLAC National Accelerator Laboratory, Menlo Park, CA 94025}
\affiliation{Department of Materials Science and Engineering, Stanford University, Stanford, CA 94305}

\author{\\Nathan J. Bittner}
\affiliation{Independent Researcher}

\author{Kenji Watanabe}
\affiliation{Research Center for Electronic and Optical Materials, National Institute for Materials Science, 1-1 Namiki, Tsukuba 305-0044, Japan}

\author{Takashi Taniguchi}
\affiliation{Research Center for Materials Nanoarchitectonics, National Institute for Materials Science,  1-1 Namiki, Tsukuba 305-0044, Japan}

\author{Bede Pittenger}
\affiliation{Bruker Nano Surfaces, AFM Unit, Santa Barbara, CA 93117}

\author{Christina J. Newcomb}
\affiliation{Stanford Nano Shared Facilities, Stanford University, Stanford, CA 94305}

\author{\\Marc A. Kastner}
\affiliation{Stanford Institute for Materials and Energy Sciences, SLAC National Accelerator Laboratory, Menlo Park, CA 94025}
\affiliation{Department of Physics, Stanford University, Stanford, CA 94305}
\affiliation{Department of Physics, Massachusetts Institute of Technology, Cambridge, MA 02139}

\author{Andrew J. Mannix}
\affiliation{Stanford Institute for Materials and Energy Sciences, SLAC National Accelerator Laboratory, Menlo Park, CA 94025}
\affiliation{Department of Materials Science and Engineering, Stanford University, Stanford, CA 94305}

\author{David Goldhaber-Gordon}
\email{goldhaber-gordon [at] stanford [dot] edu}
\affiliation{Stanford Institute for Materials and Energy Sciences, SLAC National Accelerator Laboratory, Menlo Park, CA 94025}
\affiliation{Department of Physics, Stanford University, Stanford, CA 94305}

\date{12 December 2023}

\begin{abstract}

In a stack of atomically-thin Van der Waals layers, introducing interlayer twist creates a moir\'e superlattice whose period is a function of twist angle. Changes in that twist angle of even hundredths of a degree can dramatically transform the system’s electronic properties. Setting a precise and uniform twist angle for a stack remains difficult, hence determining that twist angle and mapping its spatial variation is very important. Techniques have emerged to do this by imaging the moir\'e, but most of these require sophisticated infrastructure, time-consuming sample preparation beyond stack synthesis, or both. In this work, we show that Torsional Force Microscopy (TFM), a scanning probe technique sensitive to dynamic friction, can reveal surface and shallow subsurface structure of Van der Waals stacks on multiple length scales: the moir\'es formed between bi-layers of graphene and between graphene and hexagonal boron nitride (hBN), and also the atomic crystal lattices of graphene and hBN. In TFM, torsional motion of an AFM cantilever is monitored as it is actively driven at a torsional resonance while a feedback loop maintains contact at a set force with the sample surface. TFM works at room temperature in air, with no need for an electrical bias between the tip and the sample, making it applicable to a wide array of samples. It should enable determination of precise structural information including twist angles and strain in moir\'e superlattices and crystallographic orientation of VdW flakes to support predictable moir\'e heterostructure fabrication.

\end{abstract}

\maketitle

\section{Introduction} 
The theoretical prediction of electronic Bloch bands in moir\'e superlattices in twisted Van der Waals (VdW) bilayers ~\cite{suarez_morell_flat_2010, shallcross_quantum_2008, trambly_de_laissardiere_localization_2010, bistritzer_transport_2010, bistritzer_moire_2011} and the subsequent observations of a correlated insulator state and unconventional superconductivity in magic-angle twisted bilayer graphene (tBG)~\cite{cao_unconventional_2018, cao_correlated_2018} have unlocked a powerful new approach to tuning and discovering electronic properties of materials. tBG has displayed topological effects (orbital ferromagnetism~\cite{sharpe_emergent_2019} and quantized anomalous hall effect~\cite{serlin_intrinsic_2020}), ferroelectricity~\cite{yasuda_stacking-engineered_2021}, strange-metal behavior~\cite{cao_strange_2020, wei_strange_2023}, and more depending on interlayer twist angle, applied electric and magnetic fields, and other subtle structural features. For example, orbital ferromagnetism in tBG appears to depend on not only the twist between the two layers of graphene but also the twist between graphene and encapsulating hexagonal boron nitride (hBN) ~\cite{shi_moire_2021}. Uniaxial strain has recently been found to dramatically influence electronic properties of tBG away from magic angle ~\cite{finney_unusual_2022,wang_unusual_2022}. Beyond tBG, a burgeoning array of moir\'e systems, extending to more layers and different constituent layers, also show exciting behaviors. Unfortunately, moir\'e superlattices based on 2D materials are plagued by poor control, reproducibility, and spatial uniformity of twist angle and other structural properties ~\cite{lau_reproducibility_2022}. Convenient, rapid, and reliable techniques for imaging moir\'e superlattices will be needed to provide feedback to guide improvements in heterostructure synthesis.

Priorities for capabilities of such a technique should include: 1) imaging moir\'e superlattices on the scale of individual unit cells (ranging from nanometers to microns), 2) imaging over large areas (microns), 3) imaging subsurface moir\'e superlattices and 4) imaging atomic crystal lattices of VdW materials (sub-nanometer). This covers many but not all structural properties known to strongly influence electronic properties. As has been succinctly summarized by McGilly \textit{et al.}~\cite{mcgilly_visualization_2020}, and is still true, techniques that depend on cryogenics, ultra-high vacuum, complex infrastructure, restrictive environmental controls and/or extensive sample preprocessing (including nanofabrication) can provide powerful information but are not appropriate for quick feedback to stack synthesis. Instead we should seek a technique that is ``straightforward'': operating in air, at room temperature. To allow characterizing partially-complete stacks, the technique should not require electrical contacts or modifications to the sample or its surface, and should work on VdW stacks on soft polymers commonly used as stamps for stack assembly. Here we aim to address the need for such a rapid feedback technique.

Multiple scanning probe techniques have recently been shown to provide structural information on moir\'es. Among those, some can be used in air at room temperature, often on a commercial AFM platform, offering the promise of tight feedback for heterostructure synthesis. Conductive AFM (C-AFM) can image atomic lattices~\cite{sumaiya_true_2022}, provided an electrical contact is made to a conductive sample or a conductive substrate below an atomically-thin insulating sample. Simple tapping-mode AFM can image open-face graphene-hBN moir\'es and few-nanometer-deep hBN-hBN moir\'es with remarkable few-nanometer lateral resolution over microns ~\cite{chiodini_moire_2022}. To our knowledge this approach has not yet worked for tBG, nor has atomic-scale imaging been shown in ambient on atomically-thin stacks. Scanning Microwave Impedance Microscopy (s-MIM) has imaged open-face tBG moir\'es under ambient conditions~\cite{ohlberg_observation_2020,ohlberg_limits_2021}. Although it does not require an electrical sample contact, it does require specialized hardware and has not been shown to resolve atomic lattices. Lateral (or Friction) Force Microscopy (LFM/FFM), a varian of contact AFM focusing on lateral rather than vertical tip deflection, has perhaps come the closest to providing a facile method for mapping structural features at both moir\'e scale~\cite{zhang_dual-scale_2022, kapfer_programming_2022} and atomic lattice scale on hBN and graphite ~\cite{marsden_friction_2013, korolkov_van_2015}: evidently lateral friction forces vary with tiny changes in the positioning of the tip on the sample. Force Modulation Microscopy (FMM) and Contact Resonance AFM (CR-AFM) map topography by contact AFM while also driving the cantilever at or below a vertical (diving board) resonance of the cantilever to image the local stiffness of the surface. FMM too has been shown to image both moir\'es and atomic lattices ~\cite{adams_breakthrough_2021}. Both LFM and FMM satisfy most of the criteria laid out above but have not been shown to resolve subsurface moir\'es, to our knowledge. 

Piezoresponse force microscopy (PFM), a contact-AFM technique, has produced remarkable maps of moir\'es with few-nanometer resolution over hundreds of nanometers ~\cite{mcgilly_visualization_2020}. By superimposing two orthogonal scans, taken by rotating the sample by 90$\degree$, the full hexagonal unit cell of a tBG moir\'e has been imaged with Lateral-PFM (L-PFM). Subsurface moir\'es were also observed, though atomic lattices have not been. The authors shared their surprise that this technique would give contrast on moir\'e samples, especially tBG which lacks the inversion asymmetry necessary to generate a piezo-electric response~\cite{mcgilly_visualization_2020, moore_nanoscale_2021, bai_excitons_2020}. Though PFM is expected to require closing an electrical loop between the AFM tip and the sample, published studies suggest that PFM in fact resolves the moir\'e contrast even on insulating substrates. In attempting to replicate the beautiful maps achieved by this technique, we stumbled upon torsional resonances, sensitive to dynamic friction at the AFM tip-sample interface, as being central to resolving moir\'e contrast.

\section{Experimental}

\begin{figure*}
    \centering
    \includegraphics[width=\linewidth]{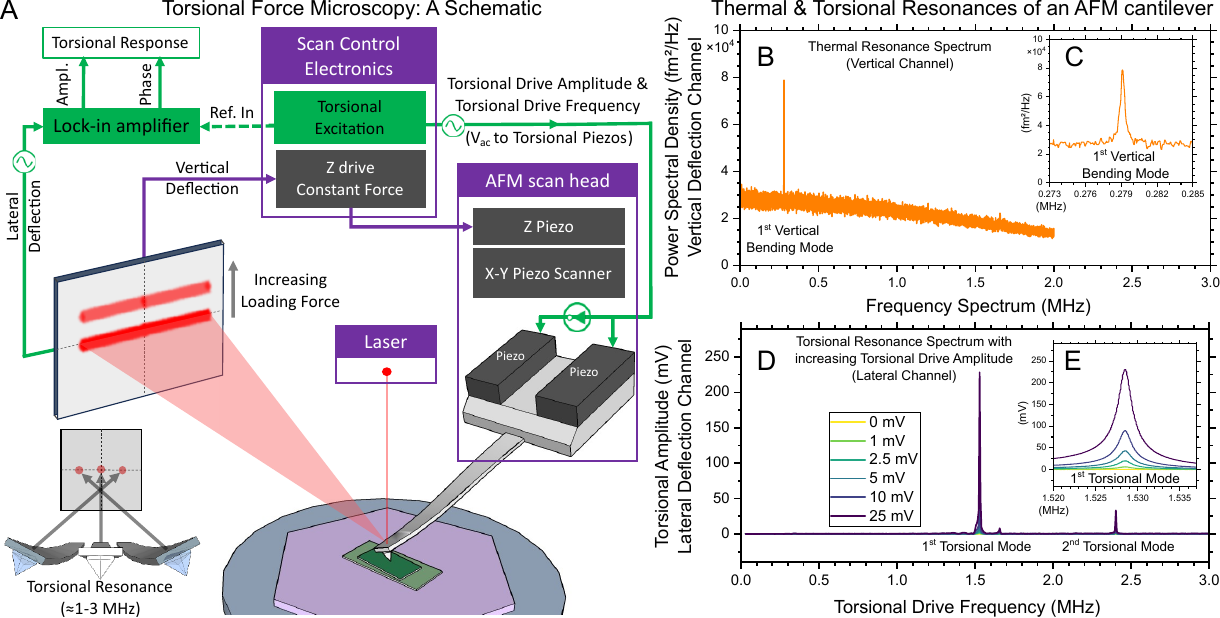}
    \caption{
        \label{Fig1}
        \textbf{Torsional Force Microscopy (TFM)} 
        (A) Schematic diagram of TFM in a system with mechanically driven torsional resonances. An incident laser beam reflects off an AFM cantilever that is driven at a torsional resonance (typically 1 to 3 MHz). This torsional resonance is mechanically excited by applying an AC drive voltage to two piezos mounted in the AFM probe holder. We measure the amplitude and phase of the resulting lateral deflection signal on the photo-detector. A constant vertical loading force between the AFM tip and the imaging surface is maintained by a feedback loop that moves the AFM tip up and down according to the topography, as in contact-AFM.
        (B) Thermally-excited resonance of an AFM cantilever (nominal spring constant 42 N/m) at room temperature in air, far from any surface, in ambient light, without any mechanical drive. The only measurable peak in power spectral density of the vertical deflection channel of the photo-detector is the fundamental resonance, i.e. the first vertical bending mode of the cantilever. (C) Zooming into a narrower frequency range shows that the resonance is at 279 kHz. (D) Torsional response of the same AFM cantilever (lateral photo-detector channel) as a function of drive frequency of the torsional piezos, for several drive amplitudes. Two prominent resonances appear at 1.529 MHz (E) and at 2.4 MHz, respectively.
        }
\end{figure*}

In this work, we map spatial variations in torsional resonances of an AFM cantilever, in a technique we term Torsional Force Microscopy (TFM). Fig.~\ref{Fig1}(A) presents a schematic diagram of the key components that enable TFM. The basic operation of TFM can be divided into two parts: first, a closed loop feedback (routed in purple arrows in Fig.~\ref{Fig1}(A)) tracks the topography to maintain a set vertical loading force; second, a torsional resonance is excited in the AFM cantilever and the mechanical response is measured in open loop (green arrows in Fig.~\ref{Fig1}(A)) The first closed feedback loop is identical to that used in contact AFM, LFM, or PFM, while the second open loop shares similarities with non-contact or tapping-mode AFM. The two loops operate in parallel. TFM does not require any electrical connections to either the tip or the sample, so the two can be electrically floating and insulating.

The bending of the cantilever as it moves into contact with the sample surface is measured as a change in the vertical position of the laser spot on a four-quadrant position sensitive photodetector. Such a photodetector provides outputs proportional to the position of the laser spot along the vertical and horizontal axes. Thermal or mechanical drift and bimetallic expansion of coated AFM tips under the incident laser led to force offsets of the order of hundreds of nanonewtons over a few hours after aligning the laser on the cantilever. This drift, if not periodically checked and corrected, can damage both the AFM tip and the sample. We developed a protocol to accurately estimate the force applied by the AFM tip on the sample surface (see supplementary materials).

In parallel to the closed feedback loop, an independent open loop maps spatial variation in the frictional response, revealing both moir\'e superlattices and atomic lattices. This open loop operates by mechanically exciting a torsional motion of the AFM cantilever, near a torsional resonance. Two piezos in the cantilever holder are driven 180$\degree$ out of phase with each other, to specifically excite torsional motion (Fig.~\ref{Fig1}(A).) This mechanical excitation of torsional resonance modes was pioneered by L. Huang \& C. Su~\cite{huang_torsional_2004, su_torsional_2007}. By sweeping torsional drive frequency, maxima in signal amplitude consistent with torsional modes of the AFM cantilever are measured (as an AC voltage) on the lateral deflection channel of the photodetector.  The amplitude of this lateral signal in volts can be used to deduce the amplitude of torsional motion of the cantilever in nanometers, in turn enabling deduction of a lateral force - orthogonal to the vertical loading force~\cite{li_lateral_2006}.

Fig.~\ref{Fig1}(B) shows the thermal resonance spectrum (without any mechanical excitation) of an AFM cantilever measured in air at room temperature, far from any surface, on the vertical deflection channel of the photodetector. The inset of Fig.~\ref{Fig1}(C) shows the first vertical bending mode of the AFM cantilever with a peak at 279 kHz. When a torsional drive is applied to the same cantilever, two resonances appear at 1.529 MHz and 2.4 MHz in the lateral deflection channel of the photodetector, as shown in Fig.~\ref{Fig1}(D). The amplitude of this response measured at the photodetector grows linearly with torsional piezo drive amplitude (Fig.~\ref{Fig1}(E)), reaching 225 mV at 25 mV torsional drive amplitude (each reported in units of zero-to-peak amplitude). For the range of cantilevers we tested, we typically see two such modes, the first between 1 and 1.6 MHz and the second between 1.4 and 3 MHz. Typically, the resonance with the highest ratio of response to drive was chosen for imaging. In the few instances when the second-most-prominent resonance was chosen, suitable results were still obtained. For a discussion on the nature of resonance modes being driven in TFM, see supplementary materials.

By calibrating the lateral deflection sensitivity of the photodetector, the torsional resonance amplitude obtained in millivolts can be associated with side-to-side deflection of the tip apex in picometers ~\cite{mullin_non-contact_2014, green_torsional_2002, green_normal_2004}. For a test AFM cantilever with a torsional resonance at 1.4 MHz, we extracted a deflection sensitivity of 3 pm/mV (or 3 nm/V), in line with values reported in literature. Such sensitivities have substantial uncertainty (perhaps 30\%), dominated by variation in height of tip relative to nominal values~\cite{mullin_non-contact_2014}. For further details, including the estimation of peak-to-peak amplitude of torsional oscillation, see supplementary materials.

A lock-in amplifier operating near the torsional drive frequency demodulates the measured torsional amplitude and phase at every pixel. Typical line scan speeds (each line consisting of both trace and retrace) ranged from 2 Hz over microns, to 4 Hz over hundreds of nanometers and up to 30 Hz over tens of nanometers. At these speeds, the lock-in amplifier input bandwidth was typically set between the lower end of 0.211 kHz (limited by electronics) to 10 kHz, with increasing bandwidth at increasing speeds, to avoid digitization. A standard operating procedure (SOP) to set up TFM is provided in the supplementary materials.

\section{Results}

\begin{figure}[b]
    \centering
    \includegraphics[width=\linewidth]{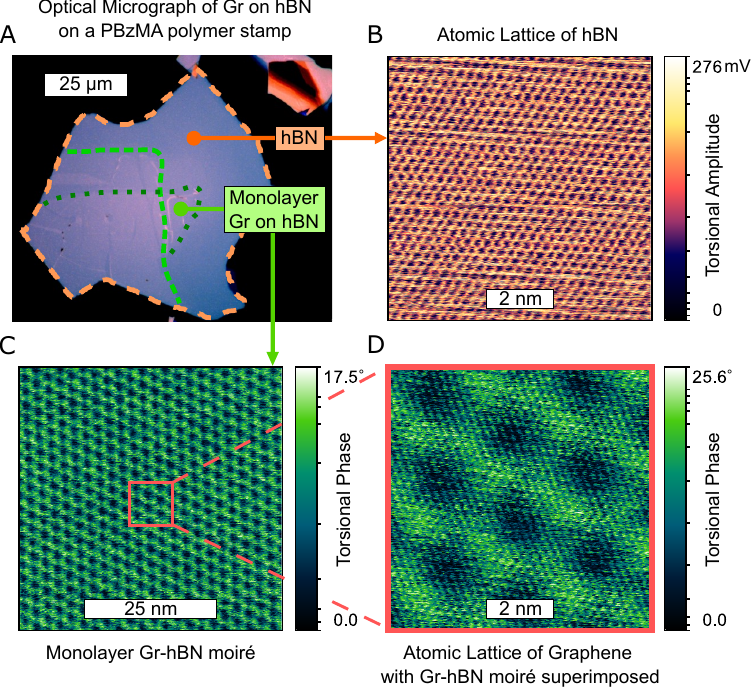}
    \caption{
        \label{Fig2}
        \textbf{Torsional Force Microscopy of the Atomic Lattice of hBN \& Graphene \& of a moir\'e of Monolayer Graphene on hBN}
        (A) Optical microscope image of a graphene-hBN heterostructure on a multilayer polymer stamp with the color scale adjusted to highlight the contrast between graphene and hBN. The dashed outlines are a guide to the eye indicating the dimensions of the two graphene flakes (green) and the hBN flake (orange). (B) TFM image taken at the approximate location of hBN marked in (A) by an orange circle, shows torsional amplitude revealing the atomic lattice of hBN, at a force of 50 nN and a torsional drive of 2.5 mV at a speed of 24.4 Hz per line. (C) Moir\'e superlattice formed between monolayer graphene and hBN, measured at the approximate location marked in (A) by a green circle. A moir\'e period of 2.6 nm indicates a relative twist between monolayer graphene and hBN of 5.4$\degree$. (D) Higher resolution image taken from the center of (C). The fine granular features of the moir\'e in (D) are likely the underlying lattice of graphene. (C,D) were imaged at a force 100 nN and a torsional drive of 5 mV at a speed of 8.14 Hz per line. (B,C,D)  were imaged with a 16x lateral signal amplifier enabled and at the 1.428 MHz torsional resonance, at a scan angle of 90$\degree$.
        }
\end{figure}

\begin{figure*}
    \centering
    \includegraphics[width=\linewidth]{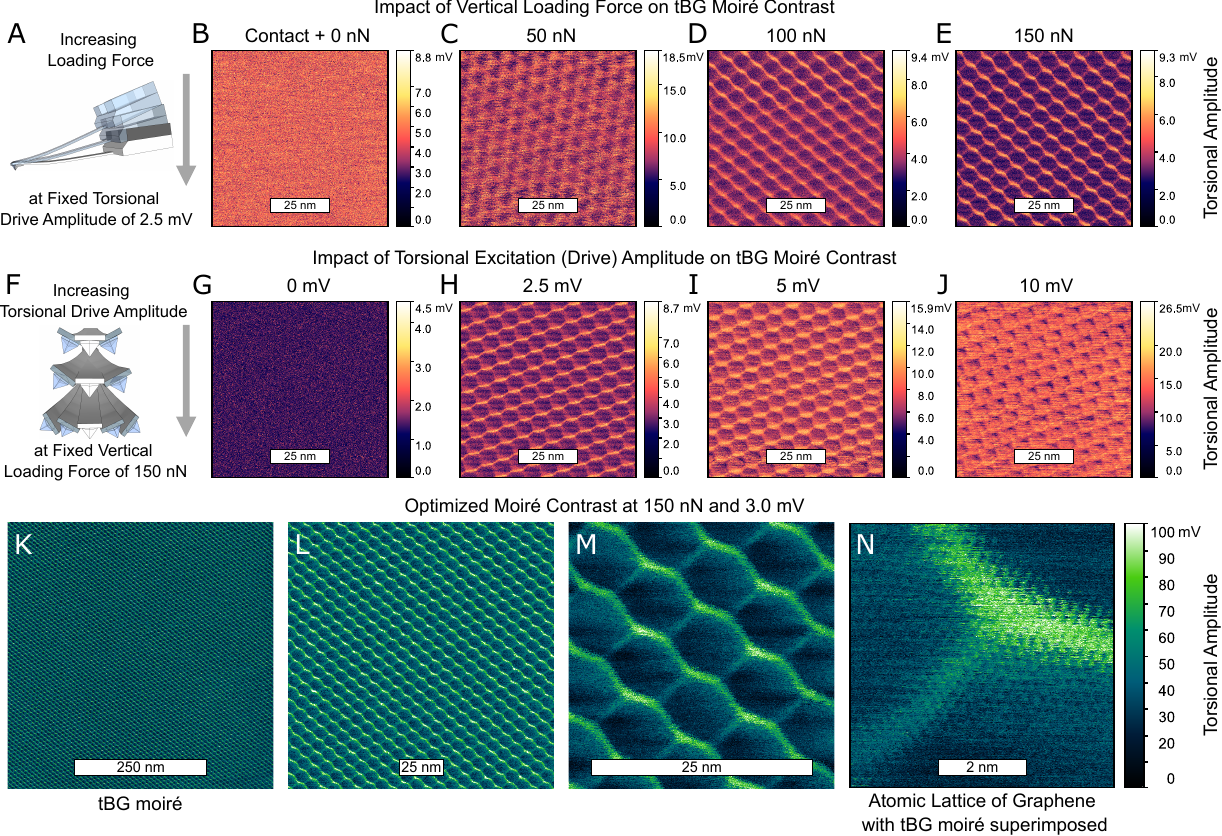}
    \caption{
        \label{Fig3}
        \textbf{Imaging moir\'es in tBG: Impact of Vertical Loading Force \& Resonant Torsional Excitation Amplitude on moir\'e contrast} 
        (A) Schematic of increasing vertical loading force at a fixed torsional excitation. (B-E) TFM maps on the surface of tBG as force is increased from the minimum required to maintain contact of the AFM tip with the sample surface (Contact + 0 nN) to 150 nN in steps of 50 nN. A moir\'e superlattice is not visible at the lowest force but becomes stronger in contrast as force is increased. (F) Schematic of increasing torsional excitation amplitude at a fixed vertical loading force. (G-J) TFM maps with torsional drive amplitude increased from 0 to 10 mV. Though no moir\'e is observed at 0 mV, a moir\'e is clearly observed at 2.5 mV. Upon further increasing the drive amplitude the moir\'e persists but the contrast in torsional amplitude decreases, instead appearing as a change in torsional phase (see supplementary Fig.S3).
        (K) Using the near-optimal imaging parameters now determined for TFM, we image a larger region, revealing a moir\'e superlattice across 500x500 nm. The moir\'e period of 7.5 nm corresponds to a twist in tBG of 1.88$\degree$. (L,M,N) Subsequent images taken at higher resolution near the center of (K). (M) shows a fine granular detail accompanying the moir\'e which is revealed in (N) to resemble an atomic lattice, most likely of the uppermost graphene surface in contact with the AFM tip, with the tBG moir\'e superimposed. 
        (B-E, G-J) were imaged at 1.4568 MHz with the lock-in amplifier bandwidth set to 102.6 kHz to ensure the resonance frequency was always within the input bandwidth. (K-N) were imaged at 1.4576 MHz, the peak of torsional resonance at 150 nN and at 3 mV drive amplitude, with the lock-in amplifier bandwidth reduced to 2 kHz and a 16x lateral signal amplifier enabled. (B-E, G-J, K-N) were all imaged at 4.07 Hz line scan speed and a scan angle of 90$\degree$.
        }
\end{figure*}

\begin{figure*}
    \centering
    \includegraphics[width=\linewidth]{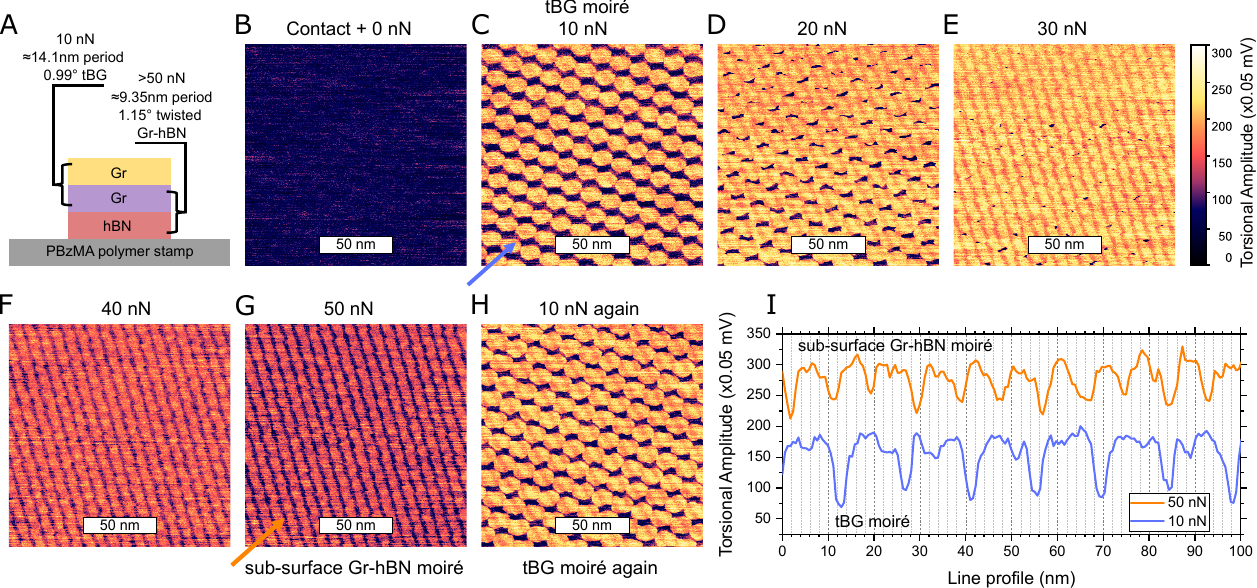}
    \caption{
        \label{Fig4}
        \textbf{Imaging a subsurface moir\'e} 
        (A) Schematic layer structure of the sample being imaged: tBG is stacked atop a flake of hBN on a multi-layer polymer stamp terminated with PBzMA. (B-H) are all imaged at the same location of the sample. (B) Resonant torsional amplitude at a vertical force equivalent to the minimum force required for the AFM tip to remain in contact with the surface (Contact + 0 nN). (C) Increasing the force to 10 nN reveals a well-defined moir\'e pattern with a period of 14.1 nm. (D-F) reveal a transition from the moir\'e observed in (C) to a moir\'e in (G) as force is stepped by 10 nN with each image. The moir\'e period observed in (G) at a force of 50 nN is 9.35 nm, different from the observed period in (C) at a force of 10 nN. (H) When force is reduced back down to 10 nN, the pattern returns to resembling that in (C), indicating the change in moir\'e from (C) to (G) is not a temporary surface cleaning effect. (I) Line profiles, taken along lines indicated by the blue (C) and orange (G) arrows. The low-force moir\'e is likely to be a tBG moir\'e. As force is increased, it is likely that the moir\'e formed by the subsurface graphene with the underlying hBN is revealed. The period of 14.1 nm of the tBG moir\'e corresponds to a twist angle of 0.99$\degree$. The period of 9.35 nm of the Gr-hBN moir\'e corresponds to a twist angle of 1.15$\degree$. These moir\'e periods are extracted from 2D FFT of the image and not their line profiles. All images in the series were acquired at the 1.3299 MHz resonance at 0.5 Hz line scan speed at a scan angle of 0$\degree$, with the 16x lateral signal amplifier enabled. The torsional piezos were driven indirectly by crosstalk in the electronics by applying 500 mV drive amplitude - which would correspond to 10-12.5 mV of torsional drive amplitude applied directly to the piezos. 10 nN and 50 nN line profiles are offset for clarity.
        }
\end{figure*}

We now employ TFM to study a common VdW heterostructure of graphene on hBN. This sample was prepared in vacuum by picking up an exfoliated flake of hBN followed by graphene. Fig.~\ref{Fig2}(A) shows an optical microscope image of this open-face heterostructure. Fig.~\ref{Fig2}(B) shows the honeycomb atomic lattice of hBN as imaged by TFM. The atomic lattice could be measured with both the first and the second torsional resonances. Remarkably, commonly-available AFM cantilevers (radius $\leq$ 25 nm) could be used without the need for sharp AFM tips, though sharp tips were preferred (radius $\leq$ 10 nm). To counter the effects of thermal drift, piezo creep, and piezo hysteresis, fast line scan speeds of between 10-30 Hz were used over square areas typically between 4 to 20 nm on a side. Extending the piezo scanning distance along the fast scan axis by 10\% beyond the edge of the frame reduced the distortion in images. It is unclear whether the features correspond to true atomic resolution versus atomic lattice resolution (i.e., spatially averaged, eggcarton-on-eggcarton tip-sample interaction)~\cite{giessibl_afms_2005}. In any case, the ability to easily visualize the atomic crystal lattice in air at room temperature in a commercially available AFM has substantial implications for guiding stacking of atomically thin materials.

Fig.~\ref{Fig2}(C) shows TFM of a moir\'e superlattice formed between monolayer graphene and hBN with a period of a mere 2.6 nm. Here and throughout this manuscript, reported moir\'e periods are extracted from 2D FFTs. The clarity of the image highlights the impressive lateral resolution of TFM. Upon further zooming into the moir\'e structure, a periodicity consistent with the atomic lattice of either graphene or hBN emerged, superimposed on the moir\'e superlattice (see Fig.~\ref{Fig2}(D)). Supplementary Fig.~\ref{FigS2} shows the complementary TFM amplitude and phase images of Fig.~\ref{Fig2}. As the AFM tip is in direct contact with graphene while taking this image, the prominent atomic lattice is likely that of graphene. However, the vertical loading is sufficient that the underlying hBN lattice might be imaged. In addition to demonstrating the success of TFM in imaging the atomic crystal lattices of hBN and graphene, this result also confirms the sensitivity of TFM to moir\'e superlattices formed at the interface of monolayer graphene and hBN.

Next, we image a moir\'e superlattice formed in tBG. To establish reproducible conditions for imaging regardless of the AFM tip used, the sample being imaged, or other variables, Fig.~\ref{Fig3} examines the impact of two key parameters of TFM: loading force and torsional drive amplitude. These two variables in turn control the tip-sample interaction. The tBG-on-hBN open-face heterostructure was prepared in vacuum, with an intended tBG twist angle of 2$\degree$. The period of the imaged moir\'e superlattice corresponds to a twist of 1.88$\degree$.

Moir\'e superlattices in tBG imaged using contact AFM techniques such as PFM and LFM have typically been reported for forces of 20-50 nN ~\cite{mcgilly_visualization_2020, kapfer_programming_2022}. To study the impact of loading force, an accurate knowledge of the force applied is necessary (especially at low forces). We develop a protocol to accurately determine the force, starting by determining the minimum force required to keep the tip in contact with the sample. We refer to this baseline as ``Contact + 0 nN''. Fig.~\ref{Fig3}(A-E) show a schematic and then TFM images acquired as increasing force from ``Contact + 0 nN'' to 150 nN in steps of 50 nN, at a constant torsional drive amplitude of 2.5 mV. Moir\'e contrast increases dramatically as force is increased. Next, the torsional excitation's drive amplitude is increased, while keeping the drive frequency and force fixed. Though some torsional excitation is necessary, high contrast in measured signal amplitude is immediately apparent at very low drive amplitude. As drive amplitude is increased, the measured signal switches from amplitude to phase. The mechanism for this remains to be studied. 

A force of 150 nN was not required on all tBG samples; moir\'e superlattices in tBG were successfully imaged at forces from 10 nN to 300 nN. Fresh AFM tips on fresh samples enabled mapping moir\'e superlattices at comparatively lower forces. With a sharp tip apex of a fresh tip, the pressure applied on the surface is likely much greater for a given force, so we speculate that the moir\'e contrast depends directly on the pressure applied, rather than the force. Samples likely accumulate a stubborn layer of adsorbates over months demanding higher forces for imaging through these layers. The tBG sample imaged in Fig.~\ref{Fig3} was prepared over five months prior to being imaged. It was mainly stored in a nitrogen drybox, but was exposed to air for days at a time on multiple occasions.

As the force was increased from zero (not in contact with the sample) to the minimum required to remain in contact (``Contact + 0 nN'') and onward to hundreds of nanonewtons, the torsional resonant modes were observed to shift to higher frequencies. The measured amplitude of the resonance also reduced with increasing force, indicating damping of the resonance \cite{su_torsional_2007}. The shift in frequency ranged from tens of Hz to tens of kHz depending on the force applied.

Once optimal parameters of force and drive amplitude were determined, the moir\'e superlattice in tBG was imaged at varying length scales. The lock-in amplifier bandwidth was also reduced to improve SNR. Fig.~\ref{Fig3}(K-N) show a tBG moir\'e with a period of 7.5 nm, imaged with successively reduced scan area is sequentially reduced without changing any other settings. The smallest-scale map (N) covers portions of three moir\'e cells; superimposed on this moir\'e pattern is an atomic-scale periodic structure, likely the atomic lattice of graphene. These results point to the versatility of TFM in imaging both moir\'e superlattices and atomic lattices on the same sample, without having to change anything more than the frame size. Supplementary Fig.~\ref{FigS3} shows TFM phase images corresponding to the amplitude images of Fig.~\ref{Fig3}.

Going forward, we continued to follow this protocol of first determining the minimum force required to remain in contact and then stepping up from ``Contact + 0 nN'' to higher forces until satisfactory moir\'e contrast is observed. 

For one sample of tBG on hBN (schematic cross-section, Fig.~\ref{Fig4}(A)), the moir\'e superlattice observed at 10 nN (Fig.~\ref{Fig4}(B)) dramatically transformed as force was stepped up to 50 nN (C-G). Upon lowering the force back to 10 nN (H), the moir\'e returned to resemble the pattern observed at 10 nN prior to the force ramp. Line profiles (I) along arrows marked in (C) and (G) show two different periods for these images, suggesting that changing the applied force allows us to select which of two superimposed moir\'es to image. 

The moir\'e seen at low applied force is likely that of tBG, whereas the moir\'e seen at high force is likely below the surface, presumably from unintentional rotational near-alignment of graphene on hBN. The period of the first moir\'e is 14.1 nm, corresponding to tBG twisted at 0.99$\degree$. This should be compared to the 1.3$\degree$ intended fabricated twist angle of the tBG. A twist relaxation of 0.3$\degree$ is often seen at these low twist angles~\cite{carr_relaxation_2018, zhang_structural_2018, wijk_relaxation_2015}. The subsurface moir\'e period of 9.35 nm corresponds to a graphene-hBN moir\'e at 1.15$\degree$ twist. These results indicate that increasing force non-destructively allows TFM to map a subsurface moir\'e. On many additional tBG samples, we have now seen a second moir\'e corresponding to an underlying hBN's near-alignment to the subsurface layer of graphene.

This measurement was performed before we understood the mechanism for TFM imaging and the measurement was set up with excitation routed to the AFM tip, as is common in modes like PFM. We later found that due to crosstalk, torsional piezos in the probe holder were driven with an excitation 2-2.5\% of the amplitude applied to the AFM tip, and that the AFM tip was disconnected from the electrical circuit. A detailed description of this issue and a comparison of the frequency spectrum in TFM mode (directly-driven torsional piezos) vs PFM mode (crosstalk-driven torsional piezos) is shown in supplementary Fig.~\ref{FigS1}.

Fig.~\ref{FigS5} shows a 2$\times$2 $\mu$m map of a tBG moir\'e with a spatially-varying period of 44-51 nm, corresponding to twist angles around 0.3$\degree$, demonstrating that TFM can image nanometer-scale moir\'es over areas relevant to typical electronic devices. The moir\'e unit cells appear hexagonal, suggesting that the surface is unreconstructed despite the small twist angle~\cite{yoo_atomic_2019, engelke_imaging_2020}, though it is possible that TFM is not sensitive to the internal structure of the moir\'e unit cell. Though we mostly studied stacks made in vacuum, we also confirmed that TFM works on tBG-hBN samples prepared in air on PC stamps (Fig.~\ref{FigS6}). 

\section{Discussion}

We now examine the origin of both moir\'e and atomic lattice contrast in TFM. In LFM, a more commonly-used technique, the AFM cantilever's lateral deflection is measured on the photodetector as the tip is dragged along the surface. In TFM the tip again rubs against the sample surface, now at a drive frequency near the MHz resonance of the cantilever, and changes in the resonant response are measured on the photodetector. By analogy we suggest that the signal on the photodetector in TFM is a measure of tip-sample friction, as in LFM. This view is supported by our observation of increasing contrast with increasing vertical tip-sample force. Moir\'e contrast originating from friction has been reported to be velocity-dependent, so the higher tip-sample velocities in TFM may provide higher contrast for imaging 2D materials~\cite{song_velocity_2022}. 

Though both the atomic lattice of VdW materials and their moir\'es have also been imaged with LFM~\cite{zhang_dual-scale_2022}, TFM offers several advantages. First, TFM adds the ability to image subsurface moir\'e superlattices. Second, like other finite-frequency techniques TFM is resilient to electronic noise outside of the lock-in amplifier's input bandwidth.
In comparison, LFM and contact AFM operate by summing the signal from DC to a few kHz (limited by a low pass filter) and are therefore strongly affected by 1/f noise. Third, TFM can work with a wide range of cantilevers, allowing applying high vertical forces where needed to enhance moir\'e visibility. LFM by contrast uses cantilevers with a very low spring constant ($\ll$ 0.5 N/m), limiting the range of vertical forces that can be applied. Lastly, we have found that TFM can yield good contrast at any scan angle relative to the long axis of the cantilever (Fig.~\ref{Fig2} and Fig.~\ref{Fig3} used 90$\degree$ scan angle, and Fig.~\ref{Fig4} used 0$\degree$) whereas LFM requires imaging at a fixed scan angle of 90$\degree$. 

Next, we compare TFM with lateral PFM at contact resonance (or CR-PFM). Here as well, TFM offers several advantages. First, TFM can image a tBG moir\'e in a single 2-dimensional scan. In contrast, CR-PFM has been reported to require superimposing two orthogonal images, with manual rotation of the sample in-between, to fully resolve the hexagonal unit cell of a large-period moir\'e superlattice. Secondly, TFM can image atomic lattices, which has not yet been reported for PFM. Lastly, TFM does not require a conducting AFM tip with bias applied between the tip and the sample, which PFM does require.

To implement TFM in instruments lacking the capability for mechanically exciting torsional resonances in an AFM cantilever, photothermal excitation of torsional resonances has recently been demonstrated to image the atomic lattice of graphite (HOPG) and to image structural features of living cancer cells, proving the versatility of the technique~\cite{eichhorn_torsional_2022, walter_probing_2023}.

Torsional Force Microscopy (TFM) is closely related to Torsional Resonance Microscopy (TRM), a technique described by L. Huang and C. Su nearly two decades ago~\cite{huang_torsional_2004, bruker_corporation_bruker_2011}. But the distinctions are important. TRM feeds back on the torsional resonance amplitude and uses the deviation in this amplitude from its setpoint to move the Z-piezo, thus varying the vertical loading force. For imaging atomically thin materials placed on soft polymer stamps, TFM allows vertical loading force to be selected and kept steady, to balance moir\'e contrast (see Fig. 3) against the risk of tearing the materials during imaging. Incorporating a phase-locked loop (PLL) to track torsional resonance frequency as it shifts due to tip-sample interaction has been demonstrated with TRM~\cite{yurtsever_frequency_2008}, and may offer advantages in TFM. 

TFM, like other ambient-based scanning probe techniques, suffers from thermal drift, piezo creep and piezo hysteresis in the lateral (X-Y) scan axis, complicating quantitative extraction of moir\'e period and thus twist angle and strain. Temperature- and humidity-controlled enclosures, as well as correcting for piezo creep and hysteresis (either actively during imaging or using sensor data post-imaging), should help reduce these errors. Additionally, with the difference between the lattice constants of hBN and graphene being within the calibration uncertainty for ambient scanning probe techniques, TFM alone cannot be used to identify which atomic lattice is being imaged \textendash prior knowledge of the structure studied, or access to other probes, is necessary. Similarly, twist relaxation as well as unintentional deviation from the fabricated twist angle in tBG on hBN stacks makes attribution of the moir\'e observed to either of the two possible moir\'e systems (tBG or Gr-hBN), challenging, especially if only one moir\'e, of period less than 14.25 nm, has been observed.

\section{Conclusion}
An open secret in the field of VdW materials is the poor success rate of most scanning probe techniques at imaging moir\'e superlattices formed in tBG \textendash a problem not shared with the moir\'es formed in Gr-hBN, which can be imaged rather easily in conventional tapping-mode AFM. Using the SOP developed for TFM we were able to find at least one moir\'e in 94\% of the 33 regions in 32 unique samples measured. Regions that did not show a moir\'e had likely relaxed to bernal stacking. Atomic lattices were observed at an even higher success rate (see supplementary note on duration and volume of study).

To summarize, we have demonstrated the use of TFM to image atomic crystal lattices and of moir\'e superlattices formed in VdW materials, in air at room temperature. Relying on dynamic friction at the tip-sample interface, with detection sensitivity enhanced by the torsional resonance of the AFM cantilever, TFM operates without the need for any electrical contacts to either the sample or the AFM tip. Thus TFM can be applied to give tight feedback on the structure of synthesized VdW stacks, helping make such synthesis more controlled. Given the strong dependence of electronic band structure on the interlayer twist angle and its spatial variation, this could have a transformative impact on fundamental and applied research on VdW materials and devices. More broadly, TFM should find an application wherever imaging of frictional properties of surfaces gives insight into local structure. For example, TFM may enable imaging biological samples at extremely low forces, a niche hard to address with LFM.

\section{Sample preparation \& AFM measurements}
Samples prepared in vacuum as part of this work used a robotic vacuum stacking tool based on one previously developed by one of the authors~\cite{mannix_robotic_2022}. Imaging was performed with the stack placed on a PBzMA terminated multi-composition polymer stamp that was also used to pick up the hBN. This polymer stamp was prepared on a PDMS handle as described in the work just cited, and was held on a clear quartz or sapphire substrate, during both stacking and imaging. Samples prepared in air as part of this work followed the technique developed by A.L. Sharpe \textit{et al.}, using a manually operated stacking tool~\cite{sharpe_emergent_2019}. Imaging was performed with the stack still on a PC polymer stamp that was used to pick up the hBN and graphene. This polymer stamp was prepared on a PDMS handle and held on a clear glass slide, both during stacking and imaging. Post-fabrication, all samples were stored in a nitrogen drybox prior to being removed for AFM measurements. 

All AFM measurements shown as part of this work were performed at Stanford university in a shared facility instrument at room temperature, in air, without any humidity or temperature control beyond the room's air-handling, on a Bruker Dimension Icon AFM equipped with NanoScope V electronics and software version 9.40 (March 2019). As a confirmation, a few test measurements were also performed outside of the facility on a Bruker Dimension Icon AFM with NanoScope 6 electronics. No modifications were made to the hardware or the software of any of the instruments to perform these measurements. A DTRCH-AM probe holder (also used for PF-TUNA or TR-TUNA), with the tip-bias wire disconnected, was used to hold AFM tips. Various AFM tips were used to measure moir\'e superlattices and atomic lattices. Adama Innovations AD-2.8-AS \& AD-2.8-SS, Oxford Instruments Asyelec.02-R2 and MikroMasch HQ:NSC18/Pt all showed good results. AFM images were analyzed in Gwyddion. A Standard Operating Procedure (SOP) for Torsional Force Microscopy, to aid in the reproduction of these results, is provided in supplementary materials.

\section{Acknowledgments}
We thank Peter De Wolf, Ravi Chandra Chintala, Senli Guo, Shuiqing Hu, Yueming Hua, Ming Ye, Marcin Walkiewicz, James R. Williams, Sultan Malik, Benjamin E. Feldman, Benjamin A. Foutty, Carlos R. Kometter, Lukas Michalek, Abhay N. Pasupathy, Cory R. Dean, M\"aelle Kapfer, Valerie Hsieh, Roman Gorbachev, and Sung Park for fruitful discussions.

\section{Funding}
Sample preparation, measurements, and analysis were supported by the US Department of Energy, Office of Science, Basic Energy Sciences, Materials Sciences and Engineering Division, under Contract DE-AC02-76SF00515. Development of tools for robotic stacking of 2D materials were supported by SLAC National Accelerator Laboratory under the Q-BALMS Laboratory Directed Research and Development funds. All AFM imaging reported here was performed at the Stanford Nano Shared Facilities (SNSF), and stamps for stacking were prepared in Stanford Nanofabrication Facility (SNF), both of which are supported by the National Science Foundation under award ECCS-2026822.  M.P. acknowledges partial support from a Stanford Q-FARM Bloch Postdoctoral Fellowship. D.G.-G. acknowledges support for supplies from the Ross M. Brown Family Foundation and from the Gordon and Betty Moore Foundation’s EPiQS Initiative through grant GBMF9460. The EPiQS initiative also supported a symposium of early career researchers which enabled feedback from the community on this work during its development. Sandia National Laboratories is a multimission laboratory managed and operated by National Technology and Engineering Solutions of Sandia, LLC., a wholly owned subsidiary of Honeywell International, Inc., for the U.S. Department of Energy’s National Nuclear Security Administration under contract DE-NA-0003525. M.H. acknowledges partial support from the National Security Agency through the Graduate Fellowship in STEM Diversity program. K.W. and T.T. acknowledge support from the JSPS KAKENHI (Grant Numbers 21H05233 and 23H02052) and World Premier International Research Center Initiative (WPI), MEXT, Japan.

\section{Data Availability}
All data, including raw AFM images, acquired from samples presented in this work are available at the Stanford Digital Repository ~\cite{pendharkar_data_2023}.

\section{Duration and Volume of Study}
A protocol that yielded the moir\'e superlattices imaged in this work was first developed in October 2022. Between November 2022 and June 2023, at least 33 regions on 32 unique samples were studied, of which 31 regions yielded at least one moir\'e (a success rate of 94\%). 10 of these regions showed signs of two moir\'es, though a second moir\'e was not searched for and analyzed on all samples. Starting at the end of March 2023, torsional piezos were directly driven, as opposed to being indirectly driven by crosstalk. Between April and June 2023 atomic lattices of hBN and graphene were searched for in at least 8 regions of 6 unique samples, each of which yielded a discernible atomic lattice.

\section{Competing Interests}
M.A.K. currently serves as a member of the Department of Energy Basic Energy Sciences Advisory Committee. Basic Energy Sciences provided funding for this work. M.A.K. also served as an independent director on the board of Bruker Corporation until May 2023. B.P. is a senior applications scientist at Bruker Nano Surfaces. All data shown were taken on a Bruker Dimension Icon AFM at Stanford University.

\section{Author Contributions}
M.P., G.Z.J., S.J.T. and J.F., with input from C.J.N., performed the initial work on AFM imaging. M.P. identified the importance of controlling the loading force and of explicitly driving the torsional resonance in open loop. B.P. identified that instrument crosstalk was the source of torsional drive during piezoresponse force microscopy (PFM) measurements, and provided supporting evidence for this. M.P. developed the protocols for imaging both the VdW moir\'es and the VdW atomic lattices shown in this work. M.P. and S.J.T. imaged all samples using the protocols developed. S.J.T. led sample preparation in vacuum with the help of M.P., N.J.B. and M.H., while A.L.S., R.V.K. and S.S.K. worked on sample preparation in air. K.W. and T.T. prepared the hBN crystals from which flakes were exfoliated for use in all samples. M.A.K., A.J.M. and D.G.-G. supervised the project. M.P. wrote the manuscript with input from all authors.

\bibliographystyle{apsrev4-2}
\bibliography{references.bib}

\beginsupplement
\justifying

\section{A comparison of TFM with other scanning probe microscopy modes}

Table S1. compares some common scanning probe techniques used to image moir\'e superlattices in twisted bilayer graphene and to image atomic lattices in VdW materials, with TFM. The techniques compared are: Lateral Force Microscopy (LFM) and Friction Force Microscopy (FFM), Piezoresponse Force Microscopy (PFM), Conductive AFM (C-AFM), Amplitude-Modulated Kelvin Probe Force Microscopy (AM-KPFM), Frequency-Modulated KPFM (FM-KPFM), Scanning-Microwave Impedance Microscopy (S-MIM), Scanning Tunneling Microscopy (STM), AC mode AFM (AC-AFM), Force Modulation Microscopy (FMM) and Contact-Resonance AFM (CR-AFM), Torsional Resonance Microscopy (TRM) and Torsional Force Microscopy (TFM).

\begin{table}[h]
\caption{\textbf{Comparison of TFM with other scanning probe microscopy modes}}
\resizebox{\textwidth}{!}{
\begin{tabular}{ c c c c c c c c }
\hline
\hline
SPM Mode        &Topography Input   & Additional & Measured Signal & Conductive & Cantilever Driven$^a$ & tBG & Atomic\\
                & (primary feedback loop) & Input/Output &  & Tip/Sample &  & Moir\'e & Lattice \\
\hline
STM             & Tunneling Current & - & Z-piezo motion & Yes & No   & Yes & Yes \\
\hline
Contact AFM     & Vertical Deflection & - & Z-piezo motion & No & No    & - & Yes \\
LFM/FFM$^b$     & Vertical Deflection & - & Lateral Deflection & No & No    & Yes & Yes \\
PFM$^c$         & Vertical Deflection & AC bias & Lateral Deflection (AC)$^d$ & Yes & No & Yes & - \\
C-AFM           & Vertical Deflection & DC bias & Current through tip & Yes & No    & - & Yes \\
S-MIM           & Vertical Deflection & RF AC bias & Reflected RF signal & Yes & No    & Yes & - \\
\hline
AC-AFM$^e$      & Vertical Deflection (AC)$^d$ & - & Z-piezo motion & No & Yes (Vertical) & - & Yes \\
AM-KPFM         & Vertical Deflection (AC) & AC/DC bias & AC/DC bias & Yes & Yes (Vertical)   & Yes & - \\
FM-KPFM         & Vertical Deflection (AC) & AC/DC bias & AC/DC bias & Yes & Yes (Vertical)  & Yes & - \\
FMM/CR-AFM      & Vertical Deflection & Driven Vertical Resonance & Vertical Deflection (AC) & No & Yes (Vertical) & Yes & Yes \\
\hline
TRM             & Torsional Deflection (AC) & - & Lateral Deflection (AC) & No & Yes (Torsional)  & - & - \\
TFM             & Vertical Deflection & Driven Torsional Resonance & Lateral Deflection (AC) & No & Yes (Torsional)  & Yes & Yes \\
\hline
\hline
\end{tabular}}
\end{table}

Notes:

\textit{a}. In this table, a cantilever is described as driven when its motion is intentionally excited either mechanically using piezos or photothermally using laser light pulsed in a focused spot at the stem of the cantilever to cause local heating. A cantilever is not considered to be driven if the cantilever motion or fluctuations are induced by ambient room temperature (as in thermal resonance). Also, a cantilever is not considered to be driven if the motion originates only after the tip makes contact with the sample, even if such motion arises because of an applied AC bias between the tip and sample.

\textit{b}. LFM requires a scan angle of 90$\degree$. Scan angle refers to the relative angle between the cantilever and the fast scan axis; an angle of 90$\degree$ means the long axis of the cantilever is perpendicular to the fast scan axis. This enables LFM to sense local changes in the static surface friction as a change in the lateral bending of the cantilever, which is detected on the lateral deflection channel of the photodetector. 

\textit{c}. tBG moir\'es have been reported to be imaged by Lateral PFM (L-PFM). Hence, in this table the measured signal for PFM is noted as the AC component of lateral deflection. More broadly, PFM works with the AC component of both vertical and lateral deflection signals.

\textit{d}. In this table, AC refers to the AC component of the signal. Where AC has not been specified, the DC component of the signal is used.

\textit{e}. In this table, AC-AFM includes non-contact and intermittent contact AFM techniques. These include Non Contact-AFM (NC-AFM), Tapping mode AFM, Frequency Modulation-AFM (FM-AFM), Amplitude Modulation-AFM (AM-AFM), and Dynamic Force Microscopy (DFM), among others. These are in contrast with DC modes like Contact AFM and LFM where the AFM tip is in continuous contact with the sample surface during imaging.

\section{Comments on the nature of the resonance modes being driven in TFM and their detection}

For a simple cantilever, a horizontal rectangular beam supported at one end,  torsional and lateral modes are generally expected at frequencies well above the first vertical bending mode. We might not expect to drive a purely lateral bending mode by torsional excitation, nor could we sense purely lateral cantilever motion on a photodetector, as the top surface deflecting the laser beam would not change its orientation. However, the rectangular beam of a typical AFM cantilever has a pyramidal tip projecting downward from near its end, so a nominally lateral bending mode would include a torsional component of motion, enabling such a mode to be both driven and detected~\cite{eichhorn_torsional_2022, su_torsional_2007, bruker_corporation_bruker_2011}. 

In our analysis we treat the driven mode as primarily torsional. If it is primarily lateral, or has a significant lateral component, the tip motion could differ significantly from our estimate, since our calibration is based on calculating tip motion from torsional reorientation of the cantilever~\cite{mullin_non-contact_2014}.

As a test, we imaged the atomic lattice of hBN with TFM and also performed a lateral deflection sensitivity calibration in the same experiment. The atomic lattice of hBN was first observed at nominally 5 nN of vertical loading force and a torsional piezo drive amplitude of 2.0 mV. At higher loading forces of 10 and 25nN, the image quality improved and the atomic lattice could also be imaged at higher torsional drive amplitudes. We now focus on the particular case of 5 nN and 2.0 mV. At these settings, the peak of the torsional resonance was measured to be at 1.456 MHz with a zero to peak amplitude of 100 mV, with a 16X signal gain amplifier enabled. The raw amplitude can hence be estimated to be 6.25 mV, prior to the 16X amplification. Since the thermal resonance was captured without the 16X amplification, 6.25 mV is the relevant point of comparison. For atomic lattice imaging, at these settings, a scan angle of 90~\degree was chosen along with a bandwidth of 10 kHz for the lock-in amplifier. A scan size of 12x12 nm with a scan speed of 12.2 Hz and 512 samples per line yielded acceptable results.

For this calibration, we used an Adama Innovations AD-2.8-SS AFM tip with a fundamental vertical (diving board) resonance at 50.6 kHz as measured by ``thermal tune''. The cantilever width and length were measured with an optical microscope, and the thickness and height of tip apex were approximated using the nominal values provided by the manufacturer. Additionally, the thermal resonance spectra were acquired for both the vertical and lateral deflection channels using the ``high speed data capture'' function for use in the analysis. The data were analyzed in a Jupyter notebook.

From this test we find the lateral deflection sensitivity to be 3 pm/mV and using the peak torsional deflection amplitude of 6.25 mV, we approximate the peak to peak spatial deflection of the tip apex to be 40 pm, during imaging with the above specified parameters.

A step by step procedure to perform this calibration has been described at the end of the SOP. Additional information and raw data have also been made available ~\cite{pendharkar_data_2023}. These include: (1) the relevant software settings and workspace for acquisition of thermal resonance spectra of the lateral and vertical deflection channels (.bag and .wks), (2) thermal resonance spectrum acquired using the high speed data capture function (with vertical deflection as Channel A and lateral deflection as Channel B) (.hsdc), (3) a ``thermal tune'' spectrum of the fundamental vertical resonance (.txt), (4) TFM image of the atomic lattice of hBN at 5 nN and 2.0 mV drive amplitude (.spm), (5) a spectrum of driven torsional resonance at 1.4 MHz, as was used for imaging (.dat) and, (6) the Jupyter notebook used to analyze the data (.ipynb and .pdf).

\clearpage
\section{AFM measurements: \\A comparison of Torsional Resonance in TFM vs Piezoresponse Force Microscopy (PFM) Modes}

\begin{figure}[ht]
    \centering
    \includegraphics[width=0.7\linewidth]{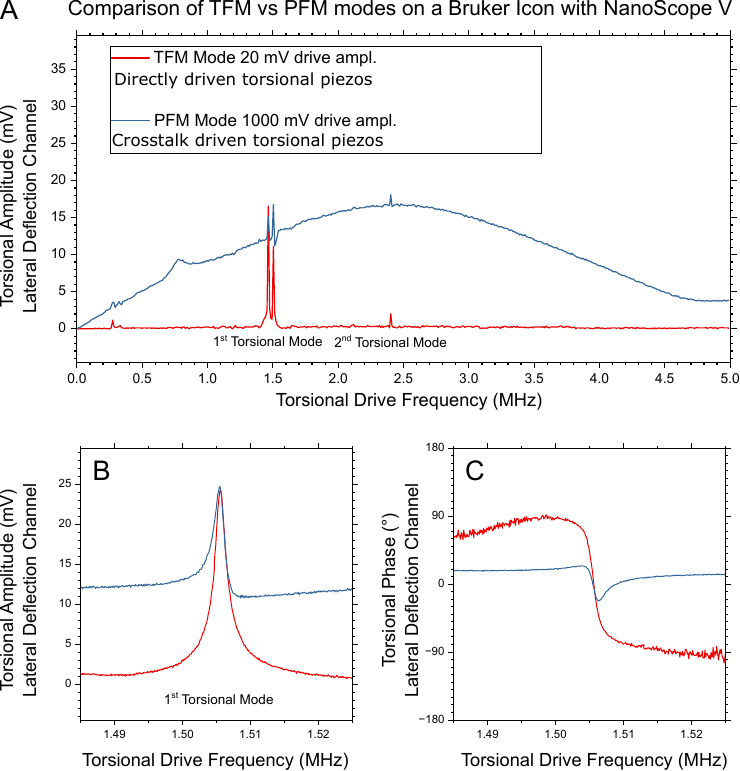}
    \caption{
        \label{FigS1}
        \textbf{Comparison of TFM vs. PFM mode of driving the torsional piezos}
        The torsional resonance spectrum from near 0 to 5 MHz is shown in the blue curve in (A) and was measured in the PFM mode by parasitically driving the torsional piezos and selecting a ``tip bias'' of 1 V. The curve shown in red in (A) shows this resonance spectrum in the TFM mode, acquired by applying 20 mV directly to the torsional piezos (applying the excitation in PFM mode required 50x higher voltage, 1V at 1.5 MHz.) (B) and (C) show a zoom-in of the first torsional mode in amplitude and phase respectively. These comparison measurements were performed in air, without any sample in proximity of the AFM tip, on an ASYELEC.02-R2 AFM tip with a nominal first vertical bending mode at 285 kHz and a nominal spring constant of 42 N/m. For the particular AFM tip chosen here, there is a split resonance  at the first torsional mode, possibly due to the properties of this particular AFM cantilever or its mechanical clamping in the probe holder. (B) and (C) zoom into one peak of that split resonance.
        }
\end{figure}

Prior to March 2023, our TFM measurements were performed by selecting piezoresponse force microscopy (PFM) mode in the Bruker Dimension Icon NanoScope software. However, because of the way we configured hardware electrical connections {\em no bias was actually applied to the tip}. Instead, through crosstalk in the electronics, torsional piezos on the probe holder were driven with amplitude of 2-2.5\% of the tip bias we selected (20-25 mV of 1V). When the drive frequency was close to a MHz torsional resonance, this drive amplitude mechanically excited torsional motion in the AFM cantilever. 
Details: The standard DTRCH-AM probe holder we used has two special features compared to most probe holders: torsional piezos and an electrically-insulating Macor AFM probe seat. In the software, we chose typical PFM settings: routing an AC bias voltage to the AFM tip, and electrically grounding the sample. However, the normal outcomes of these settings were defeated in hardware, as explained below. There are 4 electrical traces on this probe holder's PC board. A photo of the probe holder is shown in Fig.\ref{FigS7}(A), though these traces are not visible. Three of the traces connect to the torsional piezos to drive the two piezos 180$\degree$ out of phase with each other, and the fourth trace comes to a dead end on the PCB. Software configures which signal is routed through the scan head to each trace. The prominently-visible white wire bypasses the insulating probe seat to connect electrically to the tip without routing through the scan head. We verified that disconnecting this white wire and taping it to the side of the instrument (Fig.\ref{FigS7}(B)) had no effect on image contrast or the observed torsional resonance. Nor was the sample effectively grounded, as it was mounted on a polymer stamp atop a 0.5 mm thick quartz or sapphire substrate. The moir\'e image contrast we nonetheless observed when our AC drive frequency matched a torsional resonance of the cantilever was thus not from a piezoelectric response of the sample. We nonetheless refer to this as a PFM configuration because of the mode selected in software.

In PFM mode, the software-selected tip bias is routed on this probe holder to the trace that dead-ends on the PCB, while the white wire is not driven. Within the AFM scan head and the NanoScope V electronics, the wire that connects to this dead-end tip bias trace likely runs very close to the torsional piezo drive lines.  Electrical crosstalk was found to carry over about 2-2.5\% of the applied AC bias from the tip bias wire to the torsional piezos (2\% at 1.5 MHz as shown in Fig.~\ref{FigS1}(B)), so telling the system to drive the tip electrically did not in fact bias the tip, but did parasitically drive the torsional piezos.

In a newer version of the electronics (NanoScope 6), this crosstalk was found to drop to about 1\% based on a direct measurement of the voltage on the wires leading to the torsional piezos in the DTRCH-AM probe holder. 

Drawbacks of performing TFM measurements in PFM mode include: 1. Crosstalk is not an intentional part of the instrument design, so its amount may vary between instruments and probe holders (even with the same model number.) 2. The crosstalk-driven resonance in PFM mode produced an exceedingly high background signal as measured in the lateral deflection channel, leading to a poor SNR (see Fig.~\ref{FigS1}(A)\&(B)). 3. The PFM mode lacks some very useful software tools that are provided in the TR mode, which is intended for operating torsional piezos: a. option to balance the left and right piezos to optimize driving torsional resonances of a cantilever. b. option to check whether the torsional signal observed is due to torsional motion or due to unintentional coupling of vertical motion into the lateral deflection channel of the photodetector. For these reasons we have transitioned to enacting TFM explicitly in TR mode, instead of nominally performing PFM and relying on crosstalk to excite a torsional resonance.

\clearpage
\section{AFM measurements: Additional results}

\begin{figure}[ht]
    \centering
    \includegraphics[width=0.6\linewidth]{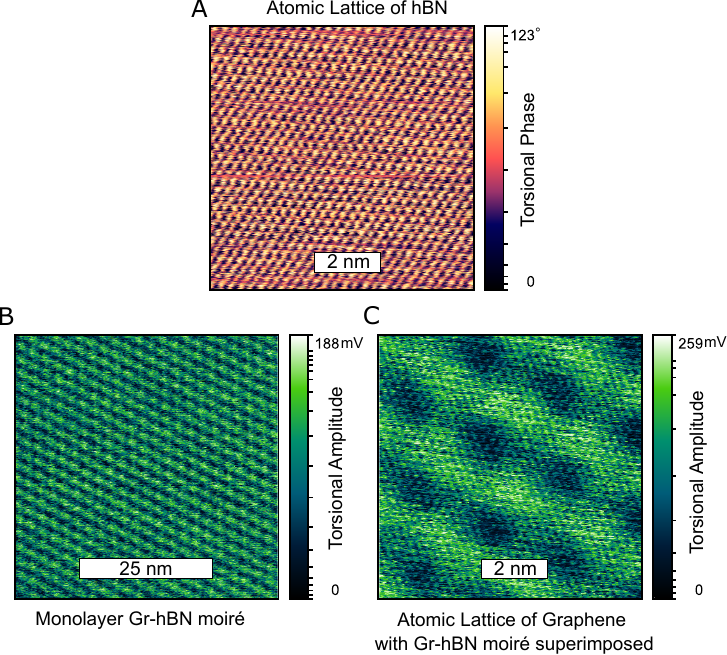}
    \caption{
        \label{FigS2}
        \textbf{Atomic Lattice of hBN \& Gr and moir\'e of monolayer Gr-hBN}
        Continuing from Fig. 2 of the main text, (A) shows the torsional phase corresponding to the torsional amplitude shown in Fig. 2 (A). (B) and (C) show the torsional amplitude of the torsional phase images shown in Fig. 2 (B) and (C). Images shown here and in Fig. 2 were processed using the align rows function in Gwyddion using polynomial fitting, followed by fixing zero to the bottom of the scale. The data was then plotted on an adaptive color scale to enhance lattice contrast. Imaging was performed using an Adama Innovations AD-2.8-SS AFM tip with a ratio of response to drive of 15 mV/mV when measured in air and away from the sample, at the 1.42 MHz resonance.
        }
\end{figure}
\begin{figure}[ht]
    \centering
    \includegraphics[width=\linewidth]{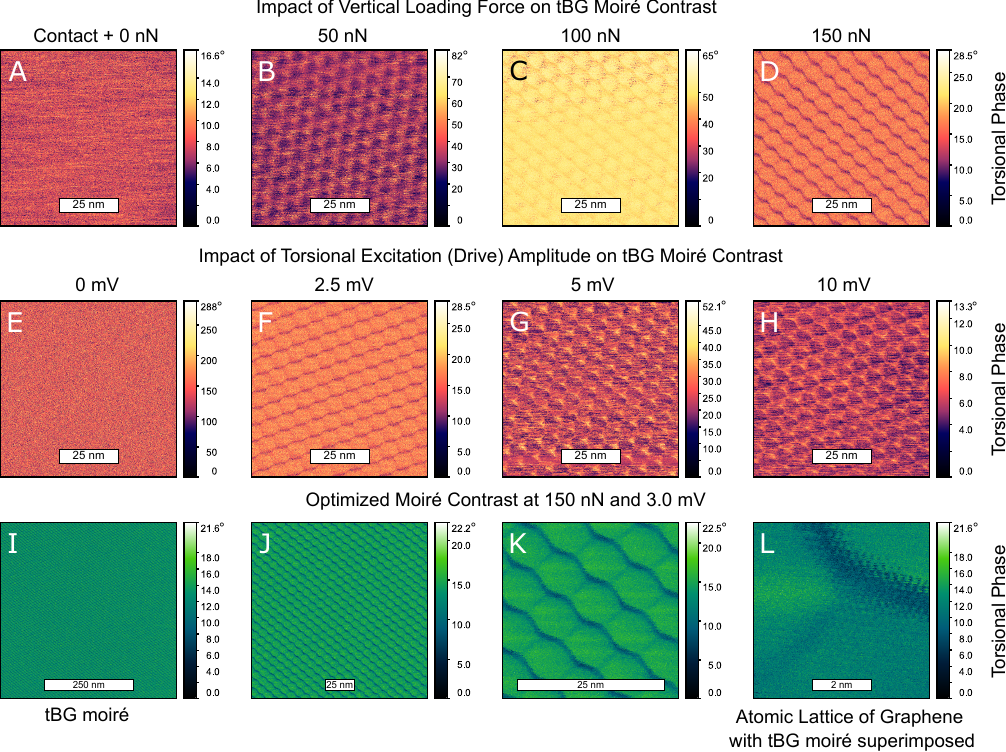}
    \caption{
        \label{FigS3}
        \textbf{Imaging moir\'es in tBG: Impact of Vertical Loading Force \& Resonant Torsional Excitation Amplitude on Moir\'e Contrast}
        Continuing from Fig. 3 of the main text, (A-L) are torsional phase counterparts of the torsional amplitude images shown in Fig. 3 of the main text. Images shown here and in Fig. 3 were processed using the align rows function in Gwyddion using polynomial fitting, followed by fixing zero to the bottom of the scale. The data were then plotted on a linear color scale. Data scale for main text Fig. 3 panels (K-N) has been restricted uniformly to 100 mV to provide a direct perspective of how the image evolves as zoomed in at the same imaging settings. Imaging was performed using an Adama Innovations AD-2.8-SS AFM tip with a ratio of photodetector response to piezo drive of 30 mV/mV when measured in air and away from the sample, at the 1.45 MHz resonance. At the time of imaging, the sample was over 5 months old and no prior surface cleaning or treatment was performed on this sample. Typically, the sample was stored in nitrogen environment with frequent removal to air for days during AFM imaging sessions.
        }
\end{figure}
\begin{figure}[ht]
    \centering
    \includegraphics[width=\linewidth]{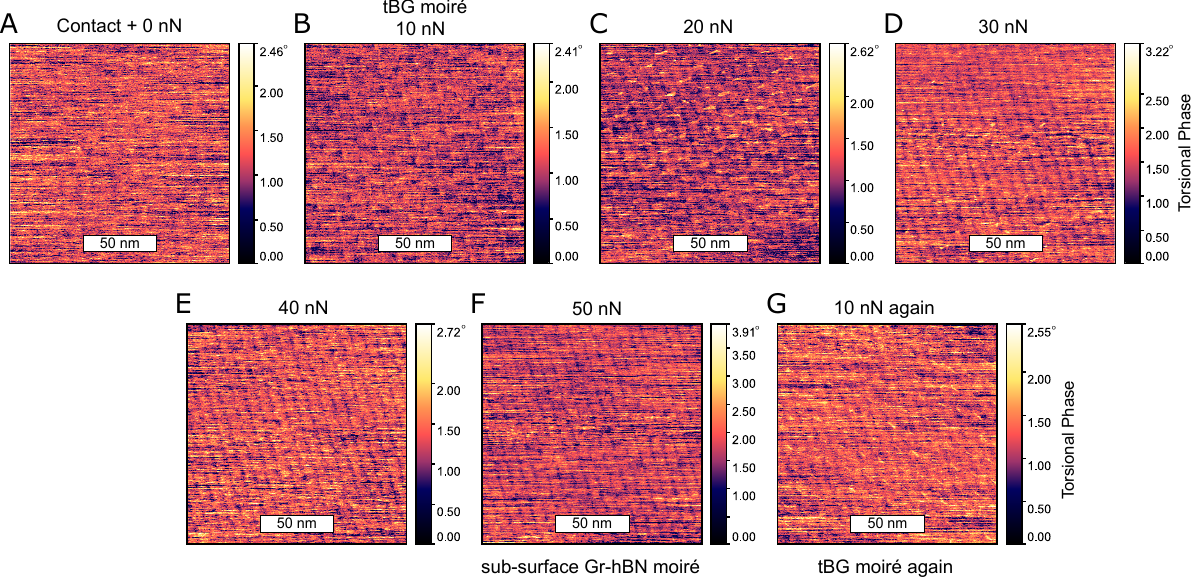}
    \caption{
        \label{FigS4}
        \textbf{Imaging a subsurface moir\'e}
        Continuing from Fig. 4 of the main text, (A-G) are torsional phase counterparts of the torsional amplitude images shown in Fig. 4 of the main text. Images shown here and in Fig. 4 were processed using the align rows function in Gwyddion using polynomial fitting, followed by fixing zero to the bottom of the scale. The data were then plotted on a linear color scale. Data scale for main text Fig. 4 panels (B-H) has been restricted uniformly to 300x0.05 mV to provide a direct perspective of how the image evolves as force is stepped at the same imaging settings. Imaging was performed using an Adama Innovations AD-2.8-AS AFM tip with a ratio of response to drive of 0.125 mV/mV when measured in air and away from the sample, at the 1.32 MHz resonance. The lower ratio of response to drive is due to the image being taken with indirectly- rather than directly-driven torsional piezos.
        }
\end{figure}

\begin{figure}[ht]
    \centering
    \includegraphics[width=0.75\linewidth]{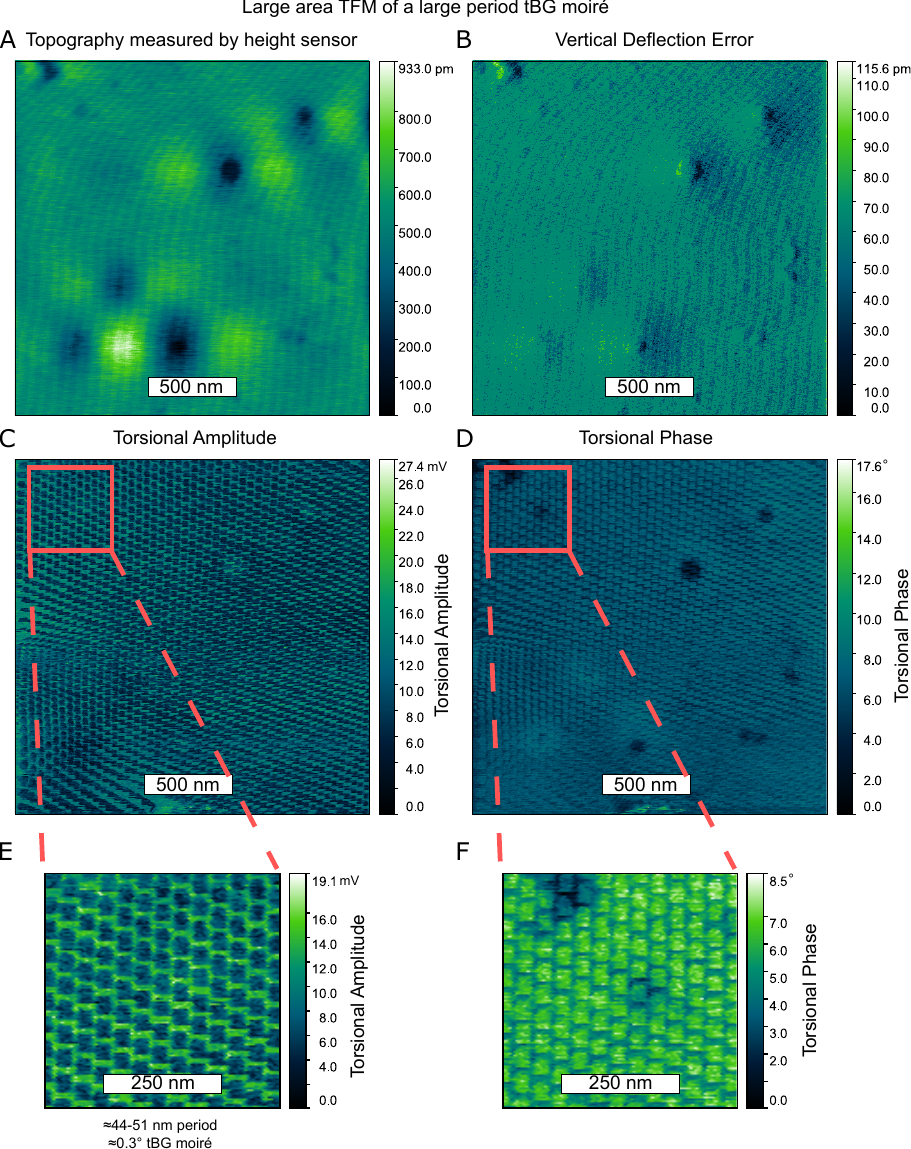}
    \caption{
        \label{FigS5}
        \textbf{Large area TFM of a large period tBG moir\'e superlattice}
        TFM was tested on a tBG stack (prepared in vacuum) over a 2x2 $\mu$ region \textendash dimensions relevant to electronic devices. (A) shows the topography measured by a height sensor and (B) shows the vertical deflection error as compared to the deflection setpoint. The faint parallel lines in (A) and (B) were determined to be periodic instrument noise. (C) and (D) show the TFM amplitude and phase, respectively. (A-D) were acquired simultaneously as separate measurement channels of the same scan. The variation of the moir\'e period over the area imaged is apparent in (C) and (D). (E) and (F) are digitally zoomed-in views of (C) and (D), respectively. The moir\'e period in (E) varies from about 44-51 nm, corresponding to a twist angle of 0.3$\degree$ in tBG. Images shown here were processed using the align rows function in Gwyddion using polynomial fitting, followed by fixing zero to the bottom of the scale. The data were then plotted on a linear color scale. The scan parameters were: Loading force of nominally 200 nN with a torsional drive amplitude of 50 mV at the 1.210 MHz torsional resonance and a line scan speed of 1.0 Hz. The scan angle was 90$\degree$ and a 16x signal amplifier was enabled. Imaging was performed using an Oxford Instruments ASYELEC.02-R2 AFM tip with a ratio of response to drive of 10 mV/mV when measured in air and away from the sample.
        }
\end{figure}

\begin{figure}[ht]
    \centering
    \includegraphics[width=0.55\linewidth]{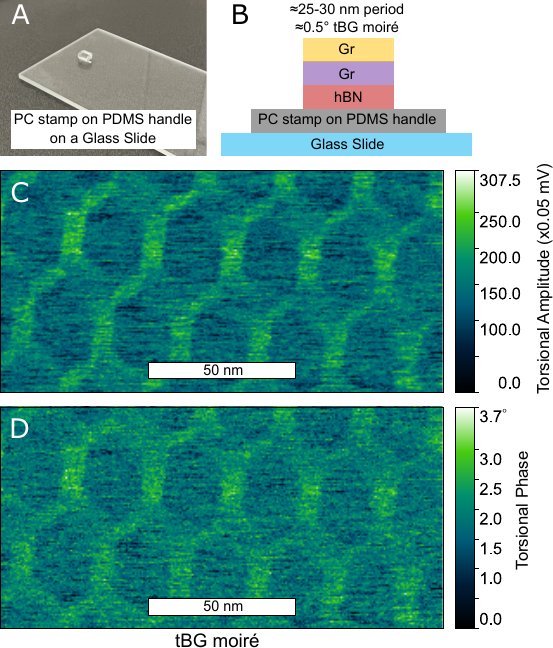}
    \caption{
        \label{FigS6}
        \textbf{TFM of tBG on a PC stamp placed on a PDMS handle on a glass slide}
        To confirm the operation of TFM on VdW heterostructures held with other common stamp surfaces, a tBG sample prepared in air on a Poly(Bisphenol A carbonate) (PC) stamp was imaged. During both stacking and imaging, this sample was held on a PDMS handle placed on a glass slide. These results show the versatility of TFM and its use in rapid feedback on stack synthesis: PC-on-PDMS-stamps held on a glass slide are commonly used to stack multiple successive VdW layers. The moir\'e period is about 25-30 nm with a computed corresponding twist angle of  0.5$\degree$. This measurement was performed with indirectly driven torsional piezos. Images shown here were processed using the align rows function in Gwyddion using polynomial fitting, followed by fixing zero to the bottom of the scale. The data were then plotted on a linear color scale. The scan parameters were: Loading force of nominally 50 nN with a 1000 mV drive amplitude at the 1.294 MHz torsional resonance and a line scan speed of 0.5 Hz. The scan angle was 0$\degree$ and a 16x signal amplifier was enabled. Imaging was performed using an Adama Innovations AD-2.8-AS AFM tip with a ratio of response to drive of 0.011 mV/mV when measured in air and away from the sample. For this particular sample, the moir\'e contrast improved as force was stepped up from ``Contact + 0 nN'' to 75 nN, at which point the high force led to a tear in the sample. The tear likely occurred due to a sharp AFM tip and a comparatively soft polymer structure (PC-on-PDMS as compared to our vacuum-compatible lithographically-patterned stamps.) When using a new type of stamp or AFM tip, force must be stepped up carefully and the effect on images examined. Once acceptable moir\'e contrast has been obtained, any further increase in force to enhance contrast must be approached with caution.
        }
\end{figure}

\clearpage
\section{Torsional Force Microscopy: a Standard Operating Procedure (SOP) \\to image VdW moir\'e superlattices and atomic lattices}
\twocolumngrid
This SOP aims to provide the necessary guidance to enable the reader to replicate the results of this work performed on a Bruker Dimension Icon AFM with NanoScope V electronics at Stanford University. It has been written assuming the reader has a basic working knowledge of operating the instrument in contact and non-contact AFM modes. This SOP is merely a suggested set of steps and not a substitute for instrument manuals, or for taking care to secure the safety of the instrument, samples, and/or users. As this protocol evolves, an updated version of this SOP may be made available ~\cite{pendharkar_data_2023}. Additional information about setting up the instrument for torsional resonance can be found elsewhere ~\cite{bruker_corporation_bruker_2011}.

\begin{figure*}
    \centering
    \includegraphics[width=0.7\linewidth]{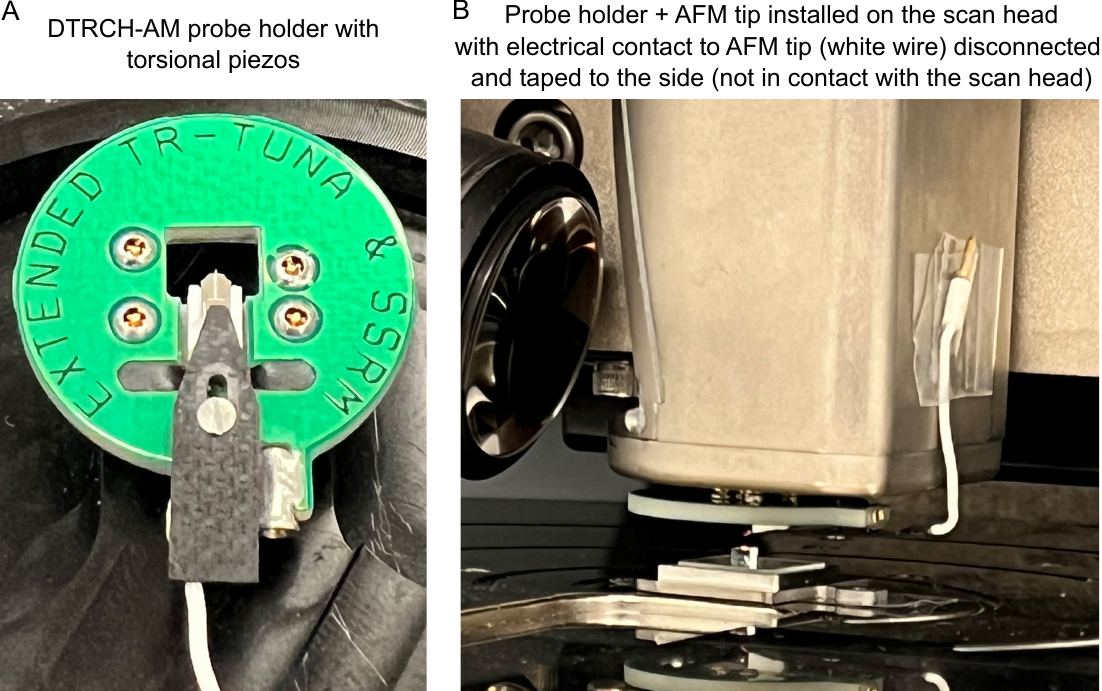}
    \caption{
        \label{FigS7}
        \textbf{Torsional probe holder}
        (A) DTRCH-AM torsional probe holder with an AFM tip inserted. The tip is seated on an electrically insulated piece of Macor, under which the two torsional piezos are housed. A clamp ( in black) holds down the AFM tip and is also the only electrical contact to the AFM tip with a white wire connected to it. (B) Probe holder installed on the scan head, with the white wire taped to the side of the scan head to ensure it doesn't make contact with the chassis.
        }
\end{figure*}

An ``ingredient list'' for TFM:
\begin{enumerate}
\item Probe holder with torsional piezos.
\begin{enumerate}
\item An AFM probe holder with torsional piezos, to mechanically excite the torsional resonance, is required for TFM. We worked with a DTRCH-AM probe holder (see Fig. S7(A)).
\item Any signal preamplification box or other associated hardware is not required for TFM and does not need to be installed.
\end{enumerate}
\item AFM tips.
\begin{enumerate}
\item Adama Innovations AD-2.8-AS \& AD-2.8-SS, Oxford Instruments Asyelec.02-R2 and MikroMasch HQ:NSC18/Pt have been tested to work, including for atomic lattice imaging.
\item AFM tips resistant to wear were preferred since some experiments required application of nominal forces exceeding 100 nN. 
\end{enumerate}
\item Sample to be imaged.
\begin{enumerate}
\item For tBG and Gr-hBN samples, a fresh sample and a fresh AFM tip were not found to be necessary for imaging a moir\'e superlattice, but moir\'e contrast was often visibly improved with both a fresh sample and a fresh tip.
\item For atomic lattice measurements of hBN, samples as old as a few years and stored in air showed a discernible atomic lattice. Atomic lattices of other materials may behave differently. Fresh samples are preferred for determination of initial imaging settings.
\end{enumerate}
\end{enumerate}

Setting up the hardware for TFM:
\begin{enumerate}
\item Put tip in probe holder \& install the probe holder.
\begin{enumerate}
\item The DTRCH-AM probe holder places the AFM cantilever on a Macor seat that electrically isolates the cantilever from the rest of the probe holder. The cantilever is secured by a clamp which has a white wire that provides the only electrical contact to the AFM tip (see Fig. S7(A)). This white wire is left disconnected and taped to the side of the scan head such that the tape ensures the metal end of the wire does not come in contact with the chassis of the AFM as shown in Fig. S7(B).
\item It may be best to center the AFM tip in the white Macor seat of the probe holder.
\item The screw on the clamp, holding down the AFM tip, should be reasonably tight. Overtightening dampens the resonance. It may take a few tries with the same AFM tip and comparing torsional resonances, to fully gauge the optimal setting.
\item We also aimed to put the AFM tip in the center of the field of view of the camera when it was fully zoomed out (no digital zoom) and aimed to align the long axis of the cantilever parallel to the camera image frame. The incident laser was found to be most focused near the center of the field of view with significant distortions at extremities. Results may vary dramatically if the laser is not focused on the AFM tip.
\item An anti-static gun for the tip and the sample may be useful but was not tested.
\item While no drawbacks are anticipated spurious voltages on scan head may affect the electrostatic potential at the tip sample interface and worse, may change during scan affecting imaging conditions.
\end{enumerate}
\item Put the sample in place \& turn on the vacuum.
\begin{enumerate}
\item[Note:] Double-sided tape to hold the sample was not tested and may not be well suited.
\end{enumerate}
\end{enumerate}

Setting the NanoScope software for TFM
\begin{enumerate}
\item Creating a Logbook. 
\begin{enumerate}
\item A table with the following columns, created in an online or offline spreadsheet, accessible post-imaging, should suffice.
\item This log will be called upon in the following steps as ``Make a logbook entry'' without further details. All applicable fields should be entered at that time and will be required for accurate estimation of force during imaging, though it would be best to enter all possible values as often as one can.
\item The log should have the following eight columns:
\begin{enumerate}
\item Time
\item Signal Sum (volts) – only an approximate value will be available after imaging begins.
\item Vertical Deflection (volts) – here, this represents the free vertical deflection of the cantilever in air, far from the sample. The value displayed in the software after imaging begins will not represent this free deflection and hence the vertical deflection column should be left blank after imaging begins. The feedback loop will try to ensure the vertical deflection is the same as the setpoint.
\item Horizontal Deflection (volts) – here, this represents the free horizontal deflection. Since there is no feedback loop that relies on this value, a drift of this deflection should not have immediate consequences to image quality. Yet, for developing an understanding of instrument drift, logging this deflection is necessary. 
\item Flying Condition (volts) – this represents the deflection setpoint, in volts, at which the tip retracts from the surface when imaging, i.e. ``flies away’’. This will be a proxy for vertical deflection during imaging albeit it may be affected by electrostatics due to the proximity of the sample surface. The tip can be considered withdrawn when the “z-piezo” indicator in the software turns red and shows the tip has moved all the way up. Recording this flying condition would require pausing imaging and reducing the voltage setpoint every 30-60 minutes to determine how much the free vertical deflection has truly drifted by.
\item Last Point of Contact (volts) – this represents the deflection setpoint at which the tip is still in contact but any further reduction of setpoint makes the tip retract (fly away) from the surface. This will be logged in conjunction with the flying condition above. This deflection setpoint will also be referred to as ``Contact + 0 nN’’ as it is the minimum force required to remain in contact with the sample in addition to any force that may be required to remain in contact.
\item Snapback (volts) – this represents the deflection setpoint at which the tip returns to making stable contact with the surface (the piezo indicator should be roughly in the middle of its range and green) after having previously retracted from the surface. Due to attractive/repulsive interactions between the tip and the sample, this deflection setpoint will differ from the last point of contact, but not by more 10s of nN.
\item Notes – enter comments like ``Laser aligned’’, ``enclosure closed’’, ``about to click engage’’, ``approached on hBN’’, ``on tBG’’, ``retracted’’, etc.
\end{enumerate}
\end{enumerate}

\item Starting ``Torsional Resonance’’ experiment.
\begin{enumerate}
\item This experiment will now configure the electronics to directly drive the torsional piezos.
\item When setting up for the first time, select ``Tapping Mode’’ followed by ``Tapping Mode in Air’’, then ``Torsional Resonance (TR-Mode)’’ and click ``Continue’’.
\begin{enumerate}
\item[Note:] Due to the number of settings that have to be changed, it is best to save the experiment once it has been correctly configured and then re-open the saved experiment every time afterwards. There is a step below to save the experiment for opening in future runs.
\end{enumerate}
\item When opening a saved experiment: 
\begin{enumerate}
\item Select Cancel on the experiment selection window that pops up.
\item From the “Experiment” menu in the row up top, select “Open experiment” and open the saved TFM experiment.
\item The steps below on configuring the experiment in ``Scan window’’ and configuring the ``Engage settings’’ can now be skipped as they will be recalled with the saved experiment. Jump to ``Align the laser on the cantilever’’.
\end{enumerate}
\end{enumerate}

\item Scan window.
\begin{enumerate}
\item With the experiment now open, it will begin with the ``Scan’’ option selected from the column on the left where ``Scan’’, ``Engage’’ and ``Withdraw’’ are show. If not, select ``Scan’’ from that column.
\item Set the file name for scans and select the user data folder.
\item In the list of scan settings, right click on white space and select “Show all” to show all the previously hidden scan settings.
\begin{enumerate}
\item With all scan settings visible, go to  ``Other’’, ``Microscope Mode’’  and select ``Dynamic Friction’’. This configures the instrument to operate in a  contact AFM like mode with a feedback loop maintaining a constant vertical deflection setpoint, irrespective of the torsional resonance settings.
\item Next, go to ``Scan’’, ``XY Closed Loop’’ and set it to ``Off’’. This will make all motions of the piezos inaccurate but is necessary for fast line scan speeds. Instead, the X-Y sensor data can be recorded as separate channels to correct for the inaccuracy, as much as possible, in postprocessing.
\item Ensure that under ``Torsion’’, ``TR Mode’’ is be set to ``Enabled’’.
\item Ensure full Z range of the fine piezos is shown in ``Limits’’, for ``Z limit’’ and ``Z range’’. This should be about 13-14 $\mu$m. If not, entering 15 (out of range) should automatically ensure these values are set to their maximum.
\item Ensure ``Amplitude Range’’ and ``TR Amplitude Range’’ are both ``4000 mV’’.
\end{enumerate}
\item Next, the ``Generic Sweep’’ button should be visible in the left most column. If it is not visible, from the options on the top of the window, select “Experiment” and configure the experiment environment to make “Generic Sweep” appear. Generic sweep will be used extensively during imaging to track the torsional resonance. 
\end{enumerate}

\item Engage Settings.
\begin{enumerate}
\item From the top of the window, select ``Microscope’’ and select ``Engage Settings’’.
\item In the ``Engage Parameters” window that pops up, right click in the white space and select ``Show All’’.
\item In ``Stage Engage’’ options, for ``SPM Engage Step’’, type in 0.02 $\mu$m. This will automatically set the minimum approach step to a value of about 0.035 $\mu$m (35 nm per step; the value of 20 nm previously entered was out of range and hence the minimum value was automatically selected). This can be increased to 100 nm if the approach is too slow. 
\item With ``Sample clearance’’ being set to 1000 $\mu$m and ``SPM safety’’ set to 100 $\mu$m, it takes about 1-2 minutes to make contact with the surface with 35 nm per step.
\item Ensure in ``Smart Engage’’ the ``Engage Mode’’ is set to ``Standard’’.
\item Do not change any other parameters. Incorrect parameters or a step size of microns in TR mode can lead to sharp tips damaging the surface of soft polymer stamps and also becoming blunt in the process.
\end{enumerate}

\item Align the laser on the cantilever
\begin{enumerate}
\item Select ``Setup’’, in the left column, to manually align the laser on the cantilever.
\begin{enumerate}
\item[Note:] Since the amplitude of torsional motion is the highest near the free end of the cantilever (on the same end as the AFM tip), the laser should be positioned as close to the free end, while still ensuring a high value of signal sum. Values typically in excess of 5 volts were common for Au coated diamond probes and in excess of 4 volts for Ti/Ir coated probes.
\item[Note:] A focused laser spot may be 40$\times$25 $\mu$m with additional lower intensity spots on either side of this ellipse.
\end{enumerate}
\item Select the correct AFM tip from the list of AFM tips.
\begin{enumerate}
\item Note the deflection sensitivity in nm/V and the nominal spring constant of the AFM tip in N/m or nN/nm
\item Thermal tune or other techniques can be used to estimate the above two parameters more accurately, if needed, but were not available in our version of the software within the torsional resonance mode.
\item Estimate the deflection in mV, per nN of force applied. It is the inverse of the value obtained by multiplying the deflection sensitivity and spring constant (nm/V $\times$ nN/nm = nN/V). For example, for a deflection sensitivity of 60 nm/V and a spring constant of 2.8 nN/nm, this comes out to about 6 mV/nN. This means that if the force has to be increased by 100 nN, the deflection setpoint must be increased (made more positive) by 600 mV.
\end{enumerate}
\item Move to the alignment station.
\item Align the laser on the AFM tip to 0 volts (0,0) on both vertical and horizontal deflection indicators.
\item Make a logbook entry.
\item[Note:] Do not move the laser or mirror deflection knobs once the laser is set as close to 0,0 as possible. The laser will heat up the cantilever and the vertical deflection value will start to drift (almost immediately) towards a more negative or positive value. This drift is expected and should reach a steady state in about 30 to 120 minutes, though the drift not reaching a steady state does not hinder measurements (logbook entries just need to be made more often, to accurately estimate the force).
\item Return from alignment station
\item Make a logbook entry.
\item[Note:] Now, it is important not to touch any of the knobs on the scan head, even if the signal keeps deviating away from 0,0 volts – we should be able to correct for the drift.
\end{enumerate}

\item Cantilever tune
\begin{enumerate}
\item With the laser aligned, click on ``Cantilever Tune’’.
\item In the tuning window that pops up, we won’t be making use of the ``Auto Tune’’ function.
\item Right click on the white space in the settings column on the right and select “Show all”.
\item The goal here is to search for a torsional resonance, and confirm that it is not a spurious vertical resonance coupling into the lateral channel (see the section on Coupling Check below) and then determine the ratio of response to drive amplitude.
\item Copy all settings as shown in Fig. S8.
\item Set the two plots to auto scale by right clicking in the plot area and selecting ``Auto scale’’.
\item Search for resonances, with a 1500 kHz center and a 3000 kHz width. At least two should show up, with a drive of 10 mV and 0.211 kHz bandwidth on the lock-in amplifier. If needed, save all spectra by clicking ``Save Curve’’.
\item The resonance spectra will look similar to plots shown in Fig. 1(D) \& (E) of the main text and Fig. S1
\item The resonance with the tallest peak can then be selected and a frequency window of about 500 kHz can set with the resonance at the center.

\begin{figure}[hb]
    \centering
    \includegraphics[width=0.75\linewidth]{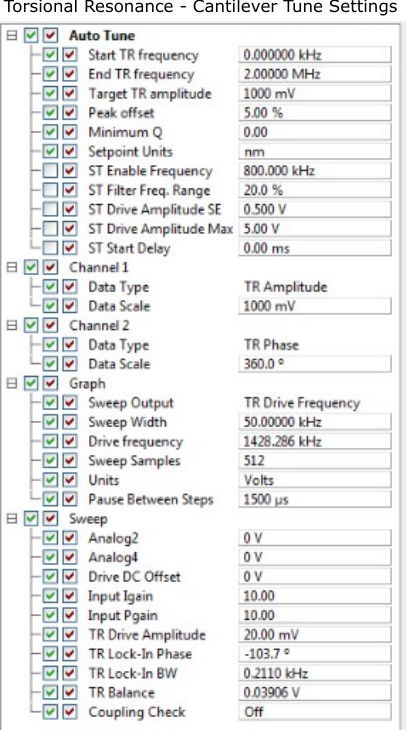}
    \caption{
        \label{FigS8}
        \textbf{Cantilever tune settings}
        The cantilever tune settings are shown here as an example and may vary between cantilevers, probe holders and AFMs. Manual tune was used to first identify resonances in the torsional frequency spectrum and then using coupling check, were confirmed not to originate from crosstalk with vertical bending modes. Then with a finer frequency sweep centered at the resonance, the piezo balance was tuned and torsional drive amplitude for initial approach was set.
        }
\end{figure}

\item Coupling check:
\begin{enumerate}
\item Next, set ``Coupling check’’ to ``On’’.
\item This mode replaces the displayed lateral amplitude with vertical amplitude, while continuing to drive the torsional resonance.
\item A purely torsional or lateral signal should not appear on the vertical deflection channel, with coupling check turned on.
\item If the resonance peak chosen appeared due to crosstalk from a vertical resonance, the previously observed peak or a shoulder of it (from the lateral channel) should become stronger. In such a case, this peak should not be chosen for imaging. If many such coupled peaks appear often, a different probe holder or different mounting of AFM tips must be tested.
\item Next, set turn off coupling check, as imaging will be performed with coupling check turned off.
\end{enumerate}

\item Balance tune.
\begin{enumerate}
\item Next, reduce the frequency window to between 50 kHz to 5 kHz with the resonance at the center.
\item Click ``More’’ at the bottom of the screen and select ``Balance Tune’’.
\item The instrument will automatically select the maxima of resonance after this. 
\item Ideally, 0V and 10V refer to driving either the left or the right piezos with 5V on balance indicating both the piezos are being driven equally. An ideal placement of an AFM tip should lead to values around 5V on balance tune showing a maxima in resonance.
\item[Note:] Software bug: As of this writing, a bug in the software limits the usability of this feature and a local maxima is observed at either 0 or 10V of balance tune, which is selected.
\item Zero the phase by clicking ``Zero phase’’.
\end{enumerate}
\item In ``Sweep’’, set ``TR drive amplitude’’ to 2 mV. A clear resonance at low drive voltages has been found to aid in imaging atomic lattices.
\item Exit cantilever tune.
\item[Note:] Since this is the first instance after aligning the laser where significant time has elapsed. The vertical deflection value should have drifted from the previously set 0V to a few hundred millivolts. The sign of this drift and the magnitude are both indicators of the drift that would have to be corrected when imaging.
\item Make a logbook entry. 
\end{enumerate}

\item Navigate to sample.
\begin{enumerate}
\item Find and focus on the sample, but aim to land on a region that is not critical as there is always a risk of the AFM tip damaging the surface if approach parameters are not chosen correctly. For a tBG/hBN open face structure, making initial contact on hBN is better than making initial contact on tBG.
\item Make a logbook entry.
\end{enumerate}

\item Check parameters:
\begin{enumerate}
\item From the settings shown in Fig. S9, copy all the settings from the ``Scan’’ and ``Feedback’’ categories. 
\item In ``Torsion’’, the value for ``TR Deflection Setpoint’’ will determine the force after contact and during imaging. It should be set to about 50 nN with respect to the ``current’’ vertical deflection value. For example, if the vertical deflection is -400 mV (negative) and the AFM tip applies 1 nN for every 6 mV of signal, applying 50 nN (+300 mV) requires that -100 mV be entered in the TR Deflection Setpoint. Forces less than 50 nN may also work depending on the AFM tips used.
\item Make a logbook entry.
\item All the other settings in the ``Torsion’’ category should have been carried over from cantilever tune, though the lock-in amplifier band width can be increased to 1 kHz to enable fast imaging initially.
\item In the ``Other’’ category, ensure the tip bias control and sample bias control are both set to ``Ground’’.
\end{enumerate}

\begin{figure}[hb]
    \centering
    \includegraphics[width=0.7\linewidth]{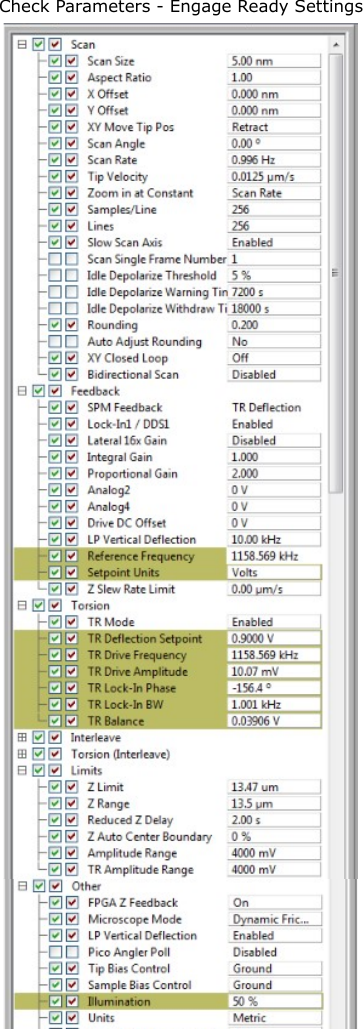}
    \caption{
        \label{FigS9}
        \textbf{Check parameters - Engage Ready Settings}
        An example of the check parameters window is shown, immediately prior to engaging with the sample. A few key checks include confirming the ``current'' vertical deflection value and appropriately choosing the ``TR Deflection Setpoint'', in volts, to make contact at about a nominal force of 50 nN.
        }
\end{figure}

\item Save the experiment and these workspace settings (to save time in future runs):
\begin{enumerate}
\item From check parameters window, go back to navigate and then to setup, to see the cantilever again.
\item Make a logbook entry.
\item Then from the Experiment menu at the top of the screen, save the experiment. This should save two files: experimentname.wks and experimentname.bag and enable recalling all the settings configured until this point the next time the instrument is turned on and this experiment started. Remember the location this experiment is saved to start from here.
\item Next, click on navigate to confirm if the region of interest is where to approach on the sample, and then click on check parameters to ensure a roughly 50 nN force is still what has been set by the TR Deflection Setpoint and the current value of vertical deflection.
\end{enumerate}

\item Engage.
\begin{enumerate}
\item With the chosen force setpoint, engage the AFM tip with the sample.
\item[Note:] It should take about a minute or two to engage. If the AFM tip is about 15 $\mu$m tall, the contact should be made with the sample at about 85 $\mu$m indicated by the current position indicator at the bottom of the screen.
\item[Note:] Keep an eye on the vertical deflection setpoint during engage. If it shifts dramatically as the tip nears the sample but while its still not in contact, it indicates an electrostatic repulsion or attraction. Approach settings would then have to be tweaked accordingly.
\end{enumerate}
\end{enumerate}

Imaging with TFM:
\begin{enumerate}
\item Setting up for imaging.
\begin{enumerate}
\item After a successful contact has been made, the instrument will start scanning the small scan region chosen entered previously.
\item The channels to be imaged would be ``Height Sensor’’, ``TR Deflection Error’’, ``TR Amplitude’’, ``TR Phase’’, ``X Sensor’’, ``Y Sensor’’. All channels should have no ``OL’’ or Off-Line plane fitting, and the ``RT’’ or Real-Time plane fitting set to ``Line’’. It may be beneficial to record all of the above channels for either trace or retrace while the remaining two channels can be TR Amplitude and TR Phase for the opposite scan direction.
\item Record movie continuously or ``forced’’ should also be enabled so that the data can now be saved.
\item To scan larger areas, increasing the P \& I settings may be required, depending upon the undulations in the sample. Values of the order of 4 (I) and 8 (P) should enable imaging over tens of microns of scan areas with 100s of nm undulation, but these may vary from instrument to instrument.
\item Next, we increase the scan area to find the region of interest (R.O.I.). This initial region of interest can be something close to the most critical R.O.I. as there is another round of optimization of force and drive frequency before we can image the most critical R.O.I. 
\item[Note:] From the main text, ``Typical line scan speeds (each line consisting of both trace and retrace) ranged from 2 Hz over microns, to 4 Hz over hundreds of nanometers and 30 Hz over tens of nanometers. At these speeds, the lock-in amplifier input bandwidth was typically set between the lower end of 0.211 kHz (limited by electronics) to 10 kHz, with increasing bandwidth at increasing speeds, to avoid digitization.’’. Adjust the scan speed and bandwidth accordingly.
\item Once the R.O.I. has been identified, reduce the scan area down to about 100 nm.
\item Then reduce the Z limit to about 4 microns. This should limit the Z range also. This option may not be available in newer versions of software and hardware.
\item Next, determine ``Contact + 0 nN’’:
\begin{enumerate}
\item Reduce the ``TR Deflection Setpoint’’, by 10 nN approximately. i.e. if the setpoint where the sample was successfully approached and the scanning is stable was -0.1 V, make it -0.15V or -0.2V. Observe the voltage on the Z piezo. 
\item At one such value the voltage on the Z piezo will become very negative, with the bar turning from green to yellow then red, indicating the piezo has retracted and is not tracking the surface anymore.
\item The goal is to determine within 10 mV the deflection setpoint at which the piezo retracts. For this, increase the setpoint again by about 0.2V to put the AFM tip back in contact and repeat the process, but with more finer steps as the ``flying condition’’ is nearing. 
\item Make a logbook entry of the flying condition and the last point of contact. 
\item Then slowly increase the force in 10 mV steps to put the tip firmly back in contact, at which point the Z piezo will remain centered in the green region and the Z voltage value shown next to it will not change as the setpoint is increased. When the AFM tip first snaps back in contact, the deflection setpoint can be logged under the ``Snapback’’ column. This value is more for information and troubleshooting and would not be necessary for imaging.
\item Make a logbook entry.
\item Next, with the ``Contact + 0 nN’’ force determined, set the value to either 10 or 20 nN, by setting the appropriate deflection setpoint.
\item This process of determining ``Contact + 0 nN’’ should be repeated every 30 minutes if the vertical drift is found to be large or every 60-120 minutes if its small. After the first few hours or imaging, the instrument should reach a steady state where this drift should become negligible over hours.
\item In the remaining text, if determine ``Contact + 0nN’’ is called, repeat the above sequence again.
\item Every time an imaging surface is changed, i.e. for example going from tBG to hBN and vice versa, this process must be repeated as tip-sample interaction can dramatically alter the force applied for nominally the same values of ``TR Deflection Setpoint’’.
\end{enumerate}
\item Next, determine the peak of torsional resonance:
\begin{enumerate}
\item During contact, due to interactions with the sample, the resonant frequency of the torsional resonance may have shifted (typically to more positive values).
\item Open the ``Generic Sweep’’ window by click on the button on the left.
\item In the window that pops up, enter 0 nm. This will ensure the AFM tip remains in contact as the sweeps are being taken. 
\item The sweep window is similar to the cantilever tune window, but now with the auto tune function removed.
\item The goal is to plot the TR Amplitude in Channel 1 and TR Phase in Channel 2 (set both scales to auto scale) and find the torsional resonance peak.
\item If a chosen peak doesn’t yield desired results, other peaks (which are confirmed to be not present in the vertical deflection channel, using coupling check) can be chosen.
\item After contact, the torsional resonance amplitude typical reduces from that in air, and the resonance frequency also shifts to a higher value. To find the peak, if it is not immediately apparent, increase the torsional drive amplitude to 10 or 20 mV and set the sweep width in the ``Graph’’ category to about 100 kHz. A peak shift of about 1 kHz may not be surprising. The shifts will greater when the force is ramped up above 100 nN to search for moir\'e superlattices.
\item Once the peak has been found, select ``Center peak’’ to center it and ``Zero phase’’. If auto centering doesn’t work, use the offset and execute buttons to manually select the peak.
\item Resonances with measured amplitudes between 10 to 50 mV would be sufficient for imaging, though optimal conditions may vary between instruments.
\item The impact of lock-in bandwidth can also be tested here. With the bandwidth set to 0.211 kHz, the noise on the resonance spectrum should be negligible. As bandwidth is ramped up, noise in both amplitude and phase should increase.
\item A bandwidth of 1 kHz may be best suited for initial imaging and can be optimized later.
\item Every time an imaging surface is changed, for optimal imaging, for example going from tBG to hBN and vice versa, this process must be repeated as tip-sample interaction can shift the torsional resonance.
\item Return to scan window by clicking ``Exit’’.
\item[Note:] Software bug: As of this writing, exiting out of the generic sweep window in the torsional resonance mode resets the lock-in amplifier bandwidth to an arbitrarily high value (between 80 to 200 kHz). This value must be immediately, manually, set to the desired values in the ``Torsion’’ category under ``TR Lock-in BW’’ field. This bug will reset the bandwidth almost every time the generic sweep window is opened and some parameters tweaked.
\end{enumerate}
\end{enumerate}

\begin{figure*}
    \centering
    \includegraphics[width=0.72\linewidth]{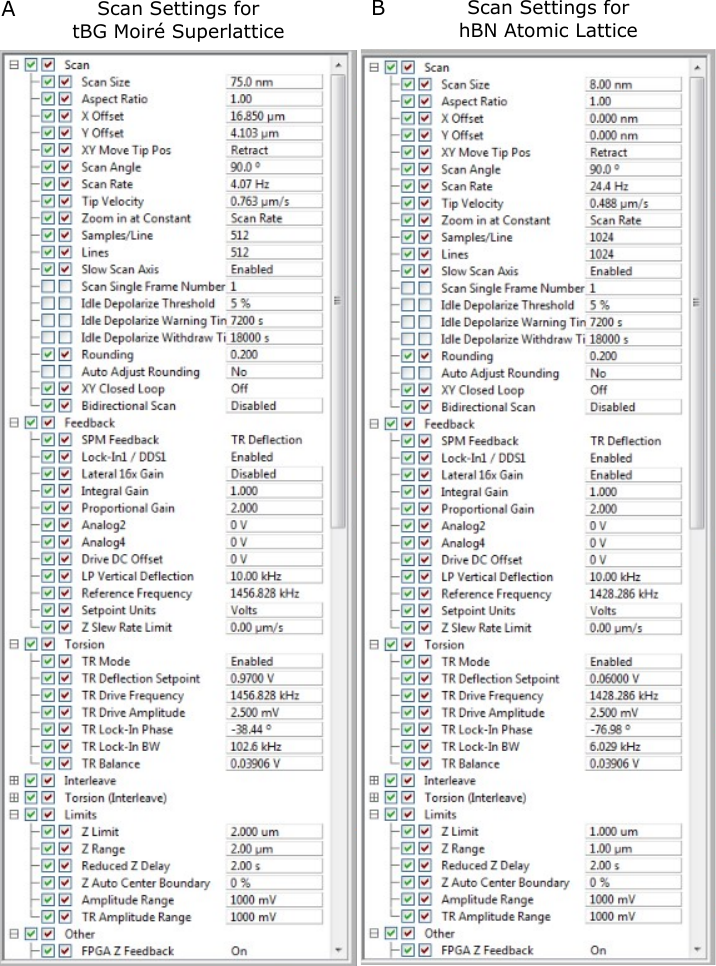}
    \caption{
        \label{FigS10}
        \textbf{Imaging settings for moir\'e superlattices and atomic lattices}
        (A) An example of settings to image a tBG moir\'e and (B) the atomic lattice of hBN are shown with the key parameters being force, shown by TR Deflection Setpoint represented in volts, torsional drive amplitude, shown by TR Drive Amplitude, in millivolts. The lock-in bandwidth for (A) is over 100 kHz as force and drive amplitude were changed to find the optimal settings for imaging and was reduced once ideal settings were found. Note the larger area image in (A) is taken at a line scan speed of 4.07 Hz while the relatively smaller image in (B) is taken at 24.4 Hz.
        }
\end{figure*}

\item Imaging moir\'e superlattices
\begin{enumerate}
\item An example of imaging conditions for moir\'e superlattices is shown in Fig. S10 (A).
\item With the above complete, move into the R.O.I. where a moir\'e superlattice is expected.
\item With the force set to Contact + 10 nN or 20 nN, image over the region of interest and zoom into the about 100 or 200 nm square. The scan speed, points per line and lock-in bandwidth can all now be increased.
\item Then, lower the force to Contact + 0 nN and image the region.
\item Step up the force in steps of 10 nN until a moir\'e appears – in either amplitude or phase or both. For example, if the most recent ``Last point of contact’’ was -0.1 V, then increase the force to -0.04V, +0.02V, +0.08V and so on. Sudden increases in force may damage both the AFM tip and sample as this is essentially a contact AFM technique.
\item Once the moir\'e contrast is optimized at a set force with a set torsional drive amplitude, open the sweep window again and tune to the peak of torsional resonance as it may have shifted with the increased force.
\item Then, step through torsional drive amplitudes to optimize contrast further. Starting from about 2.5 mV, increase the drive amplitude in steps of 2.5 mV and observe the changes to the moir\'e superlattice. A sign that the drive amplitude is too high is when sharp features of the superlattice become broadened. The lowest torsional drive that yields the desired results is preferred. Depending on the AFM cantilever, high torsional drive voltages may be needed. Due limitations of DACs, the system can only output in steps of 0.3 mV. Finer steps would not affect the output to the piezos.
\item Once both optimal conditions for force and torsional drive amplitude have been determined, open the frequency sweep again to confirm the resonance peak is selected. Then after returning from the sweep window, reduce the bandwidth to the most optimal.
\item The above protocol was used in main text Fig. 3 and has been shown. Now the scan area can be changed from many microns to many nanometers without changing anything more than scan speed and with it the bandwidth. The torsional drive amplitude and force do not have to be changed, for this region.
\item Some additional optimizations are mentioned after the note on imaging atomic lattices.
\end{enumerate}

\item Imaging atomic lattices.
\begin{enumerate}
\item An example of imaging conditions for atomic lattices is shown in Fig. S10 (B).
\item With the above complete, move into the R.O.I. where the atomic lattice has to be imaged.
\item With force set to about contact + 10 nN or contact + 20 nN, image over the region of interest and zoom into the about 5 to or 20 nm square. The scan speed, points per line and lock-in bandwidth can all now be increased.
\item Then, lower the force to Contact + 0 nN and image the region.
\item Line scan speeds of 8-30 Hz were often used in conjunction with the lowest bandwidth, for that line scan speed, that did not show any signs of digitization in the line profile.
\item Points per line was increased to 512x512 typically.
\item Step up the force in steps of 10 nN, with torsional drive amplitude set to between 2-5 mV, until an atomic lattice appears – in either amplitude or phase or both. For example, if the most recent ``Last point of contact’’ was -0.1 V, then increase the force to -0.04V, +0.02V, +0.08V and so on. Sudden increases in force may damage both the AFM tip and sample as this is essentially a contact AFM technique. Ideal contrast should not require forces much greater than 50 nN for surfaces of hBN. Other materials may behave differently. VdW flake thickness may play another critical role and thinner flakes may show poorer contrast.
\item Once the atomic lattice appears, open the sweep window again and tune to the peak of torsional resonance, again as it may have shifted with the increased force.
\item Then, step through torsional drive amplitudes to further optimize contrast. Starting from about 1 mV, increase the drive amplitude in steps of 1 mV and observe the changes to the atomic lattice. The lowest torsional drive that yields the desired results is preferred. Depending on the AFM cantilever, high torsional drive voltages may be needed. Due limitations of DACs, the system can only output in steps of 0.3 mV. Finer steps would not affect the output to the piezos.
\item Once both optimal conditions for force and torsional drive amplitude have been determined, open the frequency sweep again to confirm the resonance peak is selected. Then after returning from the sweep window, reduce the bandwidth to the most optimal.
\item The above protocol was used in main text Fig. 2(B). Now the scan area can be changed from about 3 nm to 20 nm to while keeping the speed and bandwidth the same to confirm if the image represents and atomic lattice and not periodic noise. The torsional drive amplitude and force do not have to be changed.
\item Some additional optimizations are mentioned below.
\end{enumerate}

\item Some common suggestions applicable to both moir\'e superlattice and atomic lattice imaging.
\begin{enumerate}
\item Lower the gain to value like 1(I) and 2(P), when imaging over a smooth surface. If the gain is high, ringing would be apparent in the TR Deflection Error channel.
\item The scan angle can be changed from 0$\degree$ to 90$\degree$ or other values.
\item The parameter of ``Rounding’’ set to 0.2 for scans about 100nm or smaller in open loop X-Y mode enables scanning 10\% excess on either side of the fast scan axis and may helps remove artifact at the edges of the scan frame. 
\item To optimize imaging, the ``Lateral 16x gain’’ from the feedback category can be enabled. This only amplifies the TR amplitude and not the phase.
\item Alternatively, if the signal is fairly large already  (over 100 mV), then the ``TR Amplitude Range’’ and ``Amplitude Range’’ can be reduced from 4000 mV to a lower value. At a range of 4000 mV, the lock-in is configured to take in an input of $\pm$ 2000 mV. For a 100 mV signal if 16x gain is turned on, that makes the signal 1600 mV and hence the full 4000 mV range of lock-in would be required. Conversely, if the signals are lower, the lock-in range can be reduced to increase sensitivity.
\item Remember to save the data!
\item Determine ``Contact + 0 nN’’ every 30-120 minutes to ensure only the desired force is being applied. If the drift is making the vertical deflection shift to a more negative value, the force during imaging, is unintentionally and uncontrollably increasing (even though the feedback loop is operational to maintain the deflection setpoint) and could damage the sample.
\item If both trace and retrace are being saved, the height sensor and TR deflection error channels can now be used to record the X-Y sensor data for the other pair of TR amplitude and phase.
\end{enumerate}

\item Ending the imaging session.
\begin{enumerate}
\item When the image session is coming to an end, it is important to record certain parameters so that the instrument drift (and hence drift in the force applied) can be quantified.
\item Upon clicking ``Withdraw’’, make a logbook entry.
\item Move back to check parameters and make another logbook entry. Note the presence of the sample right below the AFM tip could affect the values measured and logging these values now would aid in determining the same.
\item Next, move to navigate and move the sample to loading position.
\item Once the sample is not under the tip anymore, move to setup and make a logbook entry. Any change in value since the last recorded value is due to the presence of the sample in the vicinity. The change from when the laser was first aligned to this final logbook entry, tells the extent of drift in the force, if purely nominal force values were chosen by entering deflection in nanometers with respect to the original setpoint of 0,0V. We found that forces in excess of 100 nN can be applied, unintentionally over a course of a few hours, if not logged accurately.
\item[Note:] The experiment can again be saved, but should be saved with a different name as the conditions saved would now be for imaging and not for instrument initialization. Initializing the instrument with these settings where the Z range may have been lowered, could be dangerous.
\end{enumerate}
\end{enumerate}

Image analysis (optional)
\begin{enumerate}
\item Images presented in this work were analyzed in Gwyddion.
\item[Note:] Software bug: As of this writing, when .spm files from TFM (torsional resonance mode of the instrument) are opened in Gwyddion, the Z axis values are changed from indicating mV for amplitude and degrees ($\degree$) for phase to showing volts (V). By analyzing the Z-scale of the data in NanoScope Analysis version 3.0, we confirmed that only the displayed units of volts are incorrect and can be swapped with mV and degrees in postprocessing.
\item The data was first corrected using the align rows function followed by fixing zero to the bottom of the scale.
\item Atomic lattice images could best be represented in an adaptive color scale (due to their relatively low contrast) while moir\'e superlattice images could be represented in all color scales (typically the linear scale was chosen).
\item 2D FFT analysis was also performed using the in-built functions to measure the lattice.
\item X-Y sensor-based postprocessing to correct for scanner piezo creep and hysteresis was not employed in this work. Coupled with thermal drift, piezo creep and hysteresis introduces uncertainty in precise determination of moir\'e period, twist angle and strain as well as angular orientation of atomic lattices.
\end{enumerate}

Estimating torsional deflection sensitivity (in units of pm/mV) and amplitude of spatial torsional deflection (in units of pm, peak to peak) (optional)
\begin{enumerate}
\item[Note:] This sequence of steps assumes the reader will rely on the provided Jupyter notebook for estimation of torsional deflection sensitivity and torsional deflection amplitude. These steps have also been provided in the Jupyter notebook for ease of use.
\item Capture optical microscope images of the cantilever to be used. 
\item[Note:] If the tip height \textit{h} and thickness of the cantilever \textit{t} can be imaged with a side view or are available from SEM, then the measured \textit{h} and \textit{t} should be used.
\item Install the AFM tip in any holder (incl. the torsional probe holder). Align the laser as normal for imaging. For direct correlation with TFM results, use a torsional probe holder and perform all of the remaining steps in one sitting, i.e. without readjusting the laser spot on the AFM cantilever between calibration and TFM imaging.
\item From any imaging mode of the NanoScope software, switch the microscope mode to ``Contact'' and turn off all lock-in amplifiers in the software. Also disable the LP Friction and LP Deflection low pass filters. This should, in principle, turn off all oscillators in the system. An example of the workspace settings file to ensure all oscillators have been turned off is included in the raw data. 
\item Open HSDC capture to capture high speed data. Acquire vertical deflection and lateral deflection channel signals using the 6.25 MHz acquisition rate, for the longest allowed time (about 2.5 seconds). Save the generated .hsdc data file in a convenient location. An example .hsdc file has been included in the raw data.
\item Using thermal tune, carefully record the resonant frequency of the first vertical bending mode (not the torsional resonance frequency).
\item Import the .hsdc data in the provided Jupyter notebook (python version 3.11) and enter the cantilever dimensions and its vertical resonant frequency. In the same notebook, select the appropriate frequency range where the torsional resonance is expected (which should be the same resonance used for TFM imaging).
\item Complete any additional steps as may be mentioned in the Jupyter notebook and compute the torsional deflection sensitivity using the notebook.
\item (Optional) Image using TFM. Note the resonance frequency for imaging as well as the torsional resonance spectra used for imaging (specifically the peak torsional resonance amplitude in mV) at a set vertical loading force.
\item (Optional) From the peak of the torsional resonance measured (in mV) above, for a particular vertical loading force and imaging conditions, approximate the peak-to-peak amplitude of deflection of the tip apex, using the Jupyter notebook.
\end{enumerate}

\clearpage
\end{document}